\documentclass[12pt]{article}      
\usepackage{graphicx}

\usepackage{amsmath,amsfonts,amssymb}
\newtheorem{theorem}{Theorem}

\newtheorem{proposition}{Proposition}
\newtheorem{corollary}{Corollary}

\begin{document}

\title{Fixation probabilities for the Moran process with three or more strategies: general and coupling results
\thanks{EMF had a Capes scholarship. AGMN was partially supported by Funda\c{c}\~ao de Amparo \`a Pesquisa de Minas Gerais (FAPEMIG, Brazil).}
}
%


\author{Eliza M. Ferreira \and Armando G. M. Neves}

\author{Eliza M. Ferreira$^{1}$ \\Armando G. M. Neves$^{2}$
	\\
	\normalsize{$^{1}$Departamento de Ci\^encias Exatas, Universidade Federal de Lavras}
	\\ 	\normalsize{eliza.ferreira@dex.ufla.br}\\
	\normalsize{$^{2}$Departamento de Matem\'atica, Universidade Federal de Minas Gerais}	
	\\	\normalsize{aneves@mat.ufmg.br}
}

\maketitle

\begin{abstract}
We study fixation probabilities for the Moran stochastic process for the evolution of a population with three or more types of individuals and frequency-dependent fitnesses. Contrarily to the case of populations with two types of individuals, in which fixation probabilities may be calculated by an exact formula, here we must solve a large system of linear equations. We first show that this system always has a unique solution. Other results are upper and lower bounds for the fixation probabilities obtained by coupling the Moran process with three strategies with birth-death processes with only two strategies. We also apply our bounds to the problem of evolution of cooperation in a population with three types of individuals already studied in a deterministic setting by N\'u\~nez Rodr\'iguez and Neves (J. Math. Biol. (2016) 73:1665--1690). We argue that cooperators will be fixated in the population with probability arbitrarily close to 1 for a large region of initial conditions and large enough population sizes.

\end{abstract}

\section{Introduction}
\label{intro}

The Moran process \cite{moran} is a well known discrete-time stochastic model for the genetic evolution of a population of fixed finite size composed of individuals of several possible types (phenotypes or genotypes) and assuming no mutations in the reproduction process. The model was created having in mind a population composed of individuals of two types, assuming that the fitnesses of the individuals depend on their types and not on the frequency of these types in the population. Later on, see \cite{nowaknature} and \cite{taylor}, the process was extended to the context of Evolutionary Game Theory, see \cite{MaynardSmithPrice}, \cite{hofbauersigmund} or \cite{nowakbook}, in which fitnesses may depend on the frequency of the types among the population and are specified through a \textit{pay-off matrix}. In Evolutionary Game Theory, the types of individuals in the population are usually referred to as the \textit{strategies} adopted by the individuals.  Extension of the Moran model for populations with more than two strategies was made by \cite{wang2007evolutionary} and opens up a new class of problems, as this work will show.

In this paper we will study the Moran process for populations composed of individuals adopting three or more strategies. We acknowledge previous work by \cite{wang2007evolutionary}, but as their preprint has not been published, we will repeat part of their work here, adding full mathematical rigor. To be true, the whole paper is written having in mind that the number of strategies in the population is three, but the same theory can be easily extended to a larger number of strategies.

The first difficulty in passing from two to more than two strategies in the Moran process is that we do not have anymore an exact formula such as (\ref{piiformula}) for the fixation probabilities. Moreover, a good understanding of all possible evolutionary scenarios for the Moran process with two strategies was obtained by \cite{graphshapes} based on a classification of these scenarios provided by \cite{taylor}. As the cited works show, the behavior of the stochastic Moran process is naturally related to the behavior of the deterministic replicator dynamics \cite{taylorjonker} with the same pay-off matrix. If the number of strategies is two, we have four scenarios (without counting the trivial neutral scenario) for the replicator dynamics and eight for the Moran process \cite{taylor}. On the other hand, if the number of strategies is three, the replicator dynamics is much richer. \cite{Bomze83} showed that the number of possible phase portraits for the replicator dynamics is 47. 

In this work, after general results, we will also propose a method to produce upper and lower bounds for the fixation probabilities in the Moran process with three or more strategies. With these bounds we will understand e.g. the behavior of the fixation probability of a pure strategy which is a strict Nash equilibrium when it is close to fixation. On the other hand, we will see that in the opposite situation of a pure strategy which is a repeller in the replicator dynamics, our results are not so conclusive. Despite the progress we have made, there is still much work to be completed in the direction of providing a complete classification, as in \cite{taylor}, for the Moran process with three strategies. As an example, we will apply our results for a single phase-portrait among the 47 of \cite{Bomze83}. Our choice corresponds to one of the dynamics studied in \cite{nunezneves} for the problem of evolution of cooperation.

This paper is organized as follows. In Sect. \ref{secMorandef} we will define birth-death processes and introduce the notation for the Moran processes for two and three strategies. We will see that for three or more strategies the calculation of fixation probabilities amounts to solving a large system of linear equations. We will prove that the systems do have solutions, which are also unique. 

In Sect. \ref{seccoupling} we will present a general result for obtaining upper and lower bounds for the fixation probabilities of the Moran process with three strategies by the Markov chain coupling technique. In Sect. \ref{secnash} we will see that these bounds can produce interesting results in some situations appearing in the analysis of particular models, e.g. when we are close to a strict Nash equilibrium strategy.

Sect. \ref{seccoop} is dedicated to an example concerning the evolution of cooperation with three strategies. This example has already been studied in a deterministic setting in \cite{nunezneves}.

Finally, in Sect. \ref{secconc} we draw some conclusions and outline open problems.

Some auxiliary results on birth-death processes necessary for proving some of the Theorems on the Moran process with three strategies are proved in the Appendix.

 \section{Definitions and notations for the Moran process with three strategies}
\label{secMorandef}
\subsection{Birth-death processes}
In order to introduce the relevant concepts in this work, we start by defining \textit{birth-death processes}. As will be soon clear, the Moran process for a population with only two types of individuals is a particular case of the following definition. Although there are other different definitions, the one presented here is in \cite{nowakbook} Sect. 6.2. 

A birth-death process is a discrete-time Markov chain such that:
\begin{itemize}
	\item The set of states $S$ is finite with $S=\{0, 1, 2, \dots, N\}$.
	\item States $0$ and $N$ are absorbing, i.e. the transition probability from any of these states to a different state is null.
	\item The transition matrix of the chain is tridiagonal, i.e. from state $i\in S\setminus \{0,N\}$ the only non-vanishing transition probabilities are to state $i$ itself and to states $i \pm 1$. The transition probabilities of $i$ to $i\pm1$ are both positive. 
\end{itemize}
In the context of birth-death processes, a transition $i \rightarrow i+1$ will be termed a \textit{birth} and a transition $i \rightarrow i-1$ will be called a \textit{death}.

With the above definition, it is easy to see that all states in $S\setminus \{0,N\}$ are transient. As a consequence we have the phenomenon of \textit{fixation}: if we start at some transient state, then, if we wait long enough, with probability $1$ the state will be at either of the absorbing states. An important question is the probability that one or the other absorbing state will be attained and how this probability depends on the initial state of the population.

Due to the fact that the transition matrix is tridiagonal, the fixation probability in either absorbing state can be exactly calculated by an explicit formula known at least since \cite{moran}, see also \cite{nowakbook} for a deduction. If $X_n$ denotes the state at time $n$, let $a_i$ and $b_i$ be respectively the birth and death probabilities at state $i \in S$, i.e.
\begin{equation} \label{defbirthprob}
a_i \,=\, P(X_{n+1}=i+1 | X_n=i)
\end{equation}
and
\begin{equation} \label{defdeathprob}
b_i \,=\, P(X_{n+1}=i-1 | X_n=i) \;.
\end{equation}
Of course $a_0=b_N=0$. We let also 
\begin{equation}  \label{defri}
r_i \,=\, \frac{a_i}{b_i}\;.
\end{equation}
This latter quantity will be referred to as the the \textit{birth to death ratio}.

If $\pi_i$ denotes the probability of fixation at state $N$ when the initial state is $i$, then, of course, the probability of fixation at $0$ with the same initial state is $1-\pi_i$. Due to the fact that $0$ and $N$ are absorbing, $\pi_0=0$ and $\pi_N=1$. For the remaining values for $i$, the fixation probabilities $\pi_i$ \cite{moran}, \cite{ewens} or \cite{nowakbook}, are given by
\begin{equation}
\label{piiformula}
\pi_i \,=\, \frac{1+ \sum_{j=1}^{i-1} \prod_{k=1}^j r_k^{-1}}{1+ \sum_{j=1}^{N-1} \prod_{k=1}^j r_k^{-1}} \;,
\end{equation}
where the numerator in the right-hand side is just equal to 1 if $i=1$.

Although exact and explicit, Ewens \cite{ewens} referred to (\ref{piiformula}) as ``unwieldy", because it is difficult to qualitatively understand the sums of products in it, unless in particular simple cases. Proof of the difficulties in completely understanding birth-death and Moran processes for only two strategies is the number of papers on the subject since \cite{taylor}, e.g.  \cite{AntalScheuring}, \cite{ChalubSouza2016}, \cite{DurandLessard}, \cite{graphshapes}.

We may also interpret as deaths what we had before called births and vice-versa. This leads us to defining a \textit{dual} birth-death process as already done in \cite{graphshapes}. Let
\begin{equation} \label{defdualprocess}
\overline{a}_i= b_{N-i} \;\;\;\textrm{and}\;\;\; \overline{b}_i= a_{N-i}\;.
\end{equation}
The fixation probability at state $N$ of the dual process is calculated by a formula analogous to (\ref{piiformula}):
\begin{equation}
\label{piibarraformula}
\overline{\pi}_i \,=\, \frac{1+ \sum_{j=1}^{i-1} \prod_{k=1}^j \overline{r}_k^{-1}}{1+ \sum_{j=1}^{N-1} \prod_{k=1}^j \overline{r}_k^{-1}} \;,
\end{equation}
where $\overline{r}_k = \frac{\overline{a}_k}{\overline{b}_k}= \frac{1}{r_{N-k}}$.

Of course,
\begin{equation}
\label{recorrenciapiibarra}
\overline{\pi}_{i} = 1 - \pi_{N-i}.
\end{equation}

The duality idea is useful if we want to convert some result on the fixation probability at state $N$ in an analogous result for the fixation probability at state 0. It will be invoked again in this paper at the proof of Theorem \ref{partialexttheo}.

\subsection{The Moran process with two strategies}
Let $N$ be the fixed finite size of a population in which individuals are of two different types: either they adopt strategy A, or they adopt strategy B. The Moran process is a model for the evolution of this population in which population dynamics results from two independent random choices performed at each time step: one random individual is chosen for reproduction and one is chosen for death. We assume that the reproducing individual produces an offspring with the same type as itself. This assumption will be referred to as the \textit{absence of mutations hypothesis}. We also assume that the offspring of the reproducing individual replaces the dying individual. Population size remains thus constant in time. The dying individual is chosen uniformly among the whole population, but the reproducing individual is chosen with probability proportional to the fitness of its type in a way we will specify shortly. 

It is easy to see that, due to the absence of mutations hypothesis, the Moran process is a birth-death process as defined above, where the state $i\in S$ is the number of A individuals in the population. When the state is $i$, $a_i$ is the probability of drawing an A individual for reproduction and a B for death, and $b_i$ is the probability of drawing a B for reproduction and an A for death.

The transition probabilities $a_i$ and $b_i$ in the Moran process are calculated as follows \cite{nowaknature}. We assume that each individual interacts equally with all individuals except itself and with each interaction a ``reward'' is generated. This reward will be given by the \textit{pay-off matrix}. For a game with two-strategies, the pay-off matrix is a $2 \times 2$ matrix $M = (m_{ij})$, where $m_{ij} >0$ is the reward that an individual of type $i$ receives when interacting with an individual of type $j$. We will agree that individuals of types A and B will be labeled respectively as types $1$ and $2$ in the pay-off matrix.

The fitnesses $f_i$ of the A individuals and $g_i$ of the B individuals in general depend on the number of A and B individuals in the population and are defined \cite{nowaknature} as
\begin{equation} \label{deffi}
f_{i} \,= \, 1 - w + w \left[ m_{11}\frac{i-1}{N-1} + m_{12}\frac{N-i}{N-1}\right]
\end{equation}
and
\begin{equation}  \label{defgi}
g_{i} \,=\, 1 - w + w \left[ m_{21}\frac{i}{N-1} + m_{22}\frac{N-i-1}{N-1}\right]  \;,
\end{equation}
where the intensity of selection is $w\in [0,1]$. The larger the value of $w$, the more the game-theoretic pay-off matrix influences the fitnesses.

As already stated, the probabilities in the reproduction draw in the Moran process are proportional to the fitnesses. More exactly, the probability of drawing an A for reproduction is $\frac{if_i}{S_i}$ and the probability of choosing a B for reproduction is $\frac{(N-i)g_i}{S_i}$, where 
\begin{equation}  \label{defsi2}
S_i \,=\, i f_i+(N-i) g_i \;.
\end{equation}
As the death draw is defined to be uniform, then the probability of the transition from state $i$ to $i+1$ is
\begin{equation}  \label{aMoran2}
a_i \,=\, \frac{i f_i}{S_i} \, \frac{N-i}{N}\;.
\end{equation}
Similarly, the probability of the transition from $i$ to $i-1$ is
\begin{equation}  \label{bMoran2}
b_i \,=\, \frac{(N-i) g_i}{S_i} \, \frac{i}{N}\;.
\end{equation}
These should be substituted in (\ref{defri}), giving 
\begin{equation}
r_i\,=\, \frac{f_i}{g_i}\;,
\label{riMoran2}
\end{equation} 
which may be used in (\ref{piiformula}) for calculating the fixation probabilities for the Moran process with two strategies.

One important case in which (\ref{piiformula}) is easily understood is when all types of individuals in the population are equally fit, e.g. when all elements in the pay-off matrix are equal, or when $w=0$, a situation usually called \textit{neutral} evolution. In the neutral case, (\ref{piiformula}) yields $\pi_i=i/N$. If $\rho_A$ and $\rho_B$ are respectively the fixation probability of a single A or B individual in a population of size $N$, i.e. $\rho_A=\pi_1$ and $\rho_B=1-\pi_{N-1}$, then we have $\rho_A=\rho_B=1/N$ in the neutral case.

Another important particular case of (\ref{piiformula}) is when the fitnesses of A and B individuals are independent of the frequencies of these individuals in the population. This case is obtained by inserting $w=1$, $m_{11}=m_{12}=f$ and $m_{21}=m_{22}=g$ in (\ref{deffi}) and (\ref{defgi}), so that $f$ and $g$ become the fitnesses of A and B individuals and $r=f/g$ is the relative fitness of A individuals with respect to B. In this case the numerator and denominator in (\ref{piiformula}) become sums of finite geometric progressions with ratio $r^{-1}$ and we get
\begin{equation}
\label{freqindeppi}
\pi_i \,=\,  \frac{1-r^{-i}}{1-r^{-N}}\;.
\end{equation}

Classifying the evolutionary scenarios for the Moran process should naturally take into account the deterministic dynamics for an infinite population. The standard choice for deterministic dynamics with frequency dependent fitnesses is the replicator dynamics \cite{taylorjonker}, which inspired fitness definitions (\ref{deffi}) and (\ref{defgi}).

The classification the evolutionary scenarios for the Moran process from the point of view of fixation probabilities was performed by \cite{taylor}.  Their classification scheme considers at first whether a single A in the population is more or less fit than the Bs, and also whether a single B is more or less fit than the As. This is the natural consequence of taking into account the replicator dynamics. These fitness comparisons depend in a simple way only on the population size $N$ and on the pay-off matrix. An important discovery by \cite{taylor} is that the evolutionary scenario depends also on whether $\rho_A$ and $\rho_B$ are larger or smaller than their values $1/N$ for neutral evolution. Simple combinatorics leads to $16$ scenario possibilities, but \cite{taylor} prove that only $8$ scenarios actually exist.

The above classification for the Moran process with 2 strategies was also treated in \cite{graphshapes}. That work associates to each of the 8 evolutionary scenarios a precise shape for the graphs of the fixation probabilities. Other results in the same paper are asymptotic formulae for the fixation probabilities in the limit $N \rightarrow \infty$. As a consequence of these formulae, it can be shown that some of the evolutionary scenarios cannot happen for large enough populations. 

\subsection{The Moran process with three strategies}
Consider a fixed-size population with $N$ individuals divided into three types, say A, B and C. The state of the population at each time $n \in \{0, 1,2, \dots\}$ can be specified by the number of individuals of types A and B. Obviously, if at time $n$ we have $i$ individuals of type A and $j$ of type B, then we will have $N-i-j$ type C individuals. The state in this situation is denoted $X_n=(i,j)$. If $S=\{0,1,2, \dots, N\}$, the set of all states is
\[
\Lambda_N\,=\,\{(i,j) : i,j \in S \;\;\mathrm{and}\;\; i+j \leq N \} \;.\]

\begin{figure}
\begin{center}
\includegraphics[width= 0.5\textwidth]{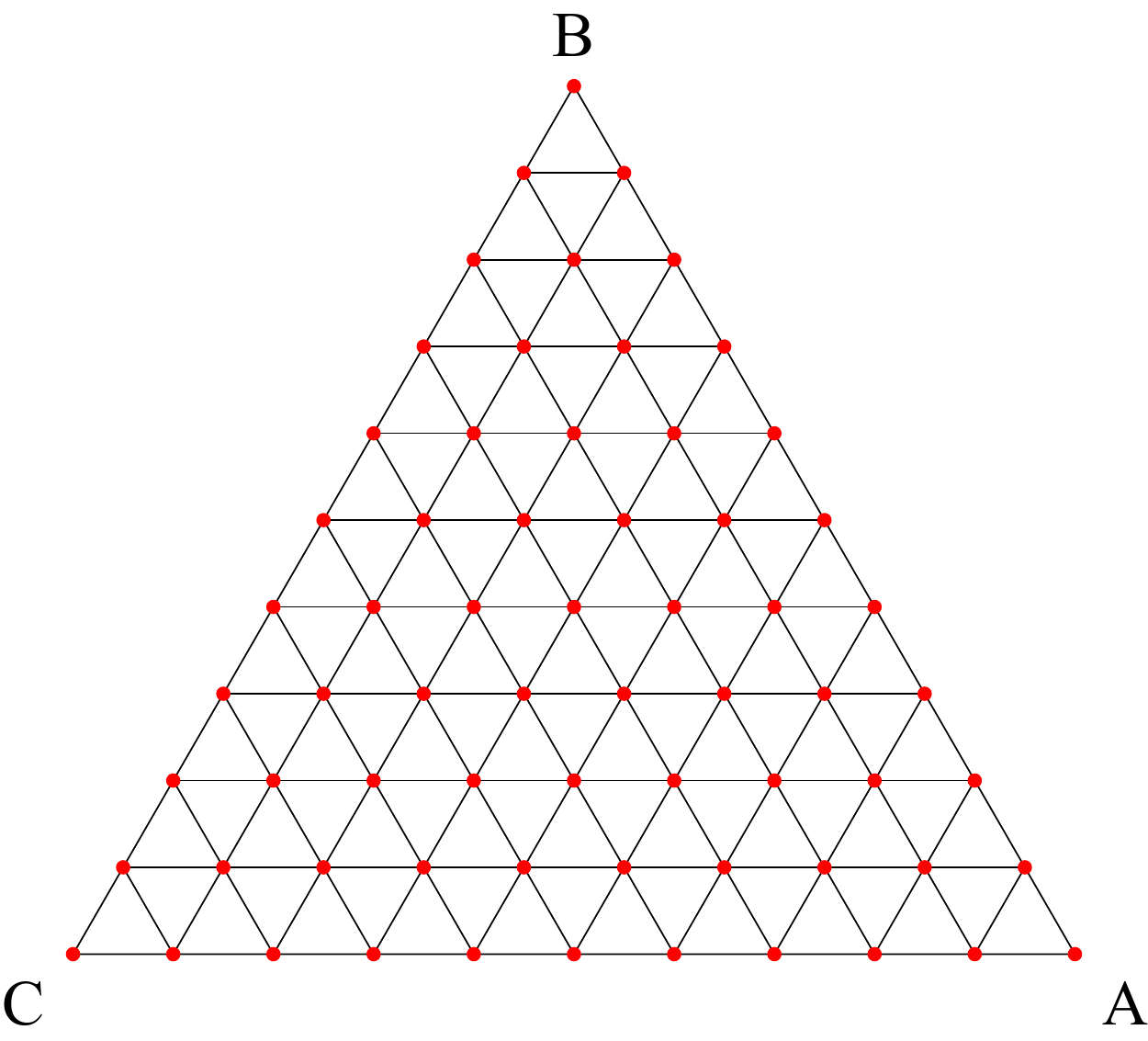}
\caption{The triangular mesh representing the states of a population in the Moran process with three strategies A, B and C. Here we have population size $N = 10$. }
\label{figrede10}
\end{center}
\end{figure}

Consider an equilateral triangle $ABC$ of unitary side length. It will be useful to represent the set of states $\Lambda_N$ of a Moran process with three strategies as the nodes on a mesh on $ABC$, see Fig. \ref{figrede10}. State $(i,j) \in \Lambda_N$ will be identified with the point on the mesh reached from the vertex C by the vector $ \frac{i}{N}\,  \vec{CA}+ \frac{j}{N} \vec{CB}$ and we will sometimes speak about states and points on the mesh as synonyms. With this identification, the vertices of the triangle represent the states in which a single type is present. On the sides of the triangle, one type is absent. 

For future use we define a compact set $\Lambda \subset \mathbb{R}^2$
by
\begin{equation}  \label{defLambda}
\Lambda=\{(x,y)\in \mathbb{R}^2: x \geq 0, y \geq 0, x+y \leq 1\} \;.
\end{equation}
We identify element $(x,y)\in \Lambda$ with the point in the triangle $ABC$ reached from the vertex C by the vector $ x\,  \vec{CA}+ y \, \vec{CB}$

As for two strategies, the population dynamics for the Moran process with three strategies is defined by two \textit{independent} random choices at each time step: one individual is drawn to die and another is drawn to reproduce, its offspring being of the same type as itself, again an \textit{absence of mutations hypothesis}. The rules are analogous to the ones already described for the Moran process with two strategies, see (\ref{defsi2}), (\ref{aMoran2}), (\ref{bMoran2}):
\begin{itemize}
	\item The death choice is made with uniform distribution.
	\item The reproduction choice is such that the probability of an individual being drawn for reproduction is proportional to the fitness of the individual's type. The exact specification will be given in the following paragraphs.
\end{itemize}

The fitness functions $f, g$ and $h$ for individuals of types respectively A, B and C may be calculated from Evolutionary Game Theory standard prescriptions. Let $M$ be a $3 \times 3$ pay-off matrix, where $m_{ij}>0$ is the reward that an individual of type $i$ receives when interacting with an individual of type $j$. We will agree that individuals of types A, B and C will be numbered, respectively, as types $1$, $2$ and $3$ in the pay-off matrix.
As natural extensions of (\ref{deffi}) and (\ref{defgi}), fitnesses are given by
\begin{eqnarray}
f_{ij} & = & 1 - w + w \left[ m_{11}\frac{i-1}{N-1} + m_{12}\frac{j}{N-1} + m_{13}\frac{N-i-j}{N-1}\right] \nonumber\\ 
g_{ij} & = & 1 - w + w \left[ m_{21}\frac{i}{N-1} + m_{22}\frac{j-1}{N-1} + m_{23}\frac{N-i-j}{N-1}\right]  \label{moran3fitnesses} \\ 
h_{ij} & = & 1 - w + w \left[ m_{31}\frac{i}{N-1} + m_{32}\frac{j}{N-1} + m_{33}\frac{N-i-j-1}{N-1}\right] \nonumber \;.
\end{eqnarray}

If the state is $(i,j)$, the probability of drawing an A for reproduction is $\frac{i f_{ij}}{S_{ij}}$, where 
\begin{equation}
\label{defsij}
S_{ij}= if_{ij} + jg_{ij} + (N-i-j)h_{ij}
\end{equation}
denotes the sum of the fitnesses of all individuals.

Analogously, the probabilities for drawing a B or a C for reproduction are respectively $\frac{j g_{ij}}{S_{ij}}$ and $\frac{(N-i-j) h_{ij}}{S_{ij}}$.

The above rules define for each $3 \times 3$  positive pay-off matrix $M$ and population size $N$ the stochastic time evolution of the population state: a discrete-time Markov chain with finite state space $\Lambda_N$. States $(0,0), (0,N)$ and $(N,0)$ are absorbing and all the remaining states are transient \cite{allen}. Again, with probability $1$, all trajectories beginning at some transient state  will be absorbed and one important problem is to calculate the fixation probability of each among the three strategies as a function of the initial state.

As the number of individuals of each type may increase or decrease by $1$ unit, or remain constant, the point representing the state either remains fixed, or jumps to one of the nearest $6$ neighbors in the triangular mesh of Fig. \ref{figrede10}. The Moran process for three strategies can thus be seen as a two-dimensional random walk on $\Lambda_N$. 

If X and Y may stand for A, B or C, we introduce now notation $p_{ij}^{XY}$ for the probability of drawing an individual of type X for reproduction and an individual of type Y for death when the state is $(i,j) \in \Lambda_N$. Transition probability $p_{ij}^{AB}$, for example, may be calculated as 
\begin{equation}  \label{pAB}
p_{ij}^{AB} = \frac{if_{ij}}{S_{ij}}\frac{j}{N}
\end{equation}
because the probability of drawing an A for reproduction is $\frac{i f_{ij}}{S_{ij}}$, the probability of drawing a B for death is $\frac{j}{N}$ and the reproduction and death draws are independent.

The other transition probabilities of the chain at state $(i,j)$, calculated in a similar way, are 
\begin{align}
p^{BA}_{ij} &= \frac{jg_{ij}}{S_{ij}}\frac{i}{N}\;, &p^{AC}_{ij} &= \frac{if_{ij}}{S_{ij}}\frac{N-i-j}{N}\;,\nonumber \\
\label{moran3prob}
p^{CA}_{ij} &= \frac{(N-i-j)h_{ij}}{S_{ij}}\frac{i}{N} \;, &p^{BC}_{ij} &= \frac{jg_{ij}}{S_{ij}}\frac{N-i-j}{N}\;, \\
p^{CB}_{ij} &= \frac{(N-i-j)h_{ij}}{S_{ij}}\frac{j}{N} \nonumber
\end{align}
and
\begin{equation} \label{moran3stay}
p^{const}_{ij}\equiv p^{AA}_{ij}+ p^{BB}_{ij}+ p^{CC}_{ij} = 1 - (p^{AB}_{ij} + p^{BA}_{ij} + p^{AC}_{ij} + p^{CA}_{ij} + p^{BC}_{ij} + p^{CB}_{ij}) \;.
\end{equation}

For the sake of future use, we observe that 
we may write equations (\ref{moran3fitnesses}) to (\ref{moran3stay}) in terms of the population fractions $x=i/N$ and $y=j/N$ of individuals of type A and B and all of them will assume the form of a term independent of $N$ plus corrections that tend to 0 when $N \rightarrow \infty$. For example, we define
the deterministic fitnesses
\begin{eqnarray}
F(x,y) & = & 1 - w + w \left[ m_{11}x + m_{12}y + m_{13}(1-x-y)\right] \nonumber\\ 
G(x,y) & = & 1 - w + w \left[ m_{21}x + m_{22}y + m_{23}(1-x-y)\right]  \label{detfitnesses} \\ 
H(x,y) & = & 1 - w + w \left[ m_{31}x + m_{32}y + m_{33}(1-x-y)\right] \nonumber \;.
\end{eqnarray}
Then
\begin{eqnarray}
f_{ij}&=& 1 - w + w \, \frac{N}{N-1}\,[ m_{11}x + m_{12}y + m_{13}(1-x-y)] \,-\, \frac{w \,m_{11}}{N-1}\nonumber\\
&=& F(x,y) \,-\, \frac{w\, m_{11}}{N-1}\,+\, w\left(\frac{N}{N-1}-1\right)[ m_{11}x + m_{12}y + m_{13}(1-x-y)]\nonumber\\
&=& F(x,y) \,+\, O(\frac{1}{N}) \label{fijlimit}\;,
\end{eqnarray}
where $x=i/N$, $y=j/N$. Analogous expressions hold for $g_{ij}$ and $h_{ij}$. In several other places in this paper we will be concerned with writing quantities depending on the state $(i,j)$ in terms of the fractions $x=i/N$ and $y=j/N$ and seeing a ``deterministic" part independent of $N$ and remainder terms that vanish as $N \rightarrow \infty$.

We can now write the equations which solutions allow us to calculate the fixation probability of individuals of types A, B or C. We will denote $\alpha_{ij}$ the fixation probability of A individuals if the initial state is $(i,j)$. For B and C individuals with the same initial state the fixation probabilities will be denoted respectively $\beta_{ij}$ and $\gamma_{ij}$. Taking into account the seven possibilities for the state after one time step and the corresponding transition probabilities to them from state $(i,j)$, the fixation probability $\alpha_{ij}$ may be written as
\begin{eqnarray} 
\alpha_{ij} & = & p^{AB}_{ij}\;\alpha_{i+1,j-1}  \; + \; p^{BA}_{ij}\;\alpha_{i-1,j+1} \; + \; p^{AC}_{ij}\;\alpha_{i+1,j} \; + \; p^{CA}_{ij}\;\alpha_{i-1,j}\; + \; p^{BC}_{ij}\;\alpha_{i,j+1}\nonumber\\ 
& & + \; p^{CB}_{ij}\;\alpha_{i,j-1} \; + \; p_{ij}^{const}\;\alpha_{ij}\;. \nonumber
\end{eqnarray}
Reorganizing the terms of the above expression we have
\begin{eqnarray} \label{0}
\displaystyle \alpha_{ij} & = & \frac{1}{1 - p^{const}_{ij}} \; \left[p^{AB}_{ij}\;\alpha_{i+1,j-1}  \; + \; p^{BA}_{ij}\;\alpha_{i-1,j+1} \; + \; p^{AC}_{ij}\;\alpha_{i+1,j} \; + \; p^{CA}_{ij}\;\alpha_{i-1,j} \right. \nonumber\\ 
& & \left. +  \; p^{BC}_{ij}\;\alpha_{i,j+1}\; 
 + \; p^{CB}_{ij}\;\alpha_{i,j-1} \right]\;.
\end{eqnarray}

Since $1 - p^{const}_{ij} = p^{AB}_{ij} + p^{BA}_{ij} + p^{AC}_{ij} + p^{CA}_{ij} + p^{BC}_{ij} + p^{CB}_{ij}$, it follows from (\ref{0}) that $\alpha_{ij}$ is a weighted average of the values of the same function in the $6$ nearest neighbors of state $(i,j)$ in the triangular mesh. The same equations remain valid if $\alpha_{ij}$ is replaced for $\beta_{ij}$ or $\gamma_{ij}$. These relations for the fixation probabilities can also be found in \cite{wang2007evolutionary}.

We have one equation of the type (\ref{0}) for each point of the mesh in the interior of the triangle $ABC$. The values of $\alpha_{ij}$ at the points of the mesh on the boundary of the triangle are known. In fact, for the points on the BC side, we have 
\begin{equation}
\alpha_{0,j} = 0 \;, \label{0boundcond}
\end{equation}
$j = 0, \dots, N$, because on this side there are no A type individuals and the absence of mutations hypothesis prohibits them to be produced by Bs or Cs. For the rest of the border we have at most two types of individuals and we may use (\ref{piiformula}). On the AB side of the triangle we thus have
\begin{equation}
\alpha_{i,N-i} = \frac{1 + \sum_{j=1}^{i-1}\prod_{k=1}^{j}(r_{k}^{AB})^{-1}}{1 + \sum_{j=1}^{N-1}\prod_{k=1}^{j}(r_{k}^{AB})^{-1}}\;,
\label{1} 
\end{equation}
$i=1, 2, \dots, N$, where $r_{k}^{AB} = \frac{f_{k,N-k}}{g_{k,N-k}}$, is the relative fitness of type A individuals with respect to B individuals in the absence of type C individuals. On the AC side of the triangle, 
\begin{equation}
\alpha_{i,0} = \frac{1 + \sum_{j=1}^{i-1}\prod_{k=1}^{j}(r_{k}^{AC})^{-1}}{1 + \sum_{j=1}^{N-1}\prod_{k=1}^{j}(r_{k}^{AC})^{-1}}\;,\label{2}
\end{equation}
for $i=1, 2, \dots, N$ and with $r_{k}^{AC} = \frac{f_{k,0}}{h_{k,0}}$.

The calculation of $\alpha_{ij}$ for the mesh points in the interior of the triangle amounts thus to solving a system of linear equations (\ref{0}) with $(N-1)(N-2)/2$ unknowns, one for each point of the mesh in the interior of the triangle. Equations (\ref{0boundcond}), (\ref{1}) and (\ref{2}) act as Dirichlet boundary conditions. This problem bears some similarities with the problem of approximating the solution of a two-dimensional Dirichlet problem for the Laplace equation using finite differences, see e.g. \cite{petrovsky}. In the approximation of the two-dimensional Laplace equation in a rectangular lattice, a simple arithmetic mean of 4 neighboring lattice points appears. In our problem, instead, we have a weighted average of 6 neighbors. 

The problems of calculating the fixation probabilities $\beta_{ij}$ for B individuals and $\gamma_{ij}$ for C individuals are completely analogous. As the transition probabilities appearing in (\ref{0}) are the same regardless we are calculating the fixation probability for A, B or C, equations (\ref{0}) are exactly the same, but boundary conditions (\ref{0boundcond}), (\ref{1}) and (\ref{2}) must be replaced by their analogues.

We will end this section by proving that the problems of finding the fixation probabilities for the Moran process with three strategies are all well-posed in the sense that the corresponding linear systems have unique solutions. The following proofs are adaptations of the corresponding ones for the finite-differences Laplace equation, see e.g. \cite{petrovsky}.

We say that $R\subset \Lambda_N$ is a connected region if for any states $(i_1,j_1)$ and $(i_2,j_2)$ in $R$ there is a path between the corresponding points in the triangle $ABC$ passing only through the links of the mesh and without passing by points representing states not in $R$. We also say that a state is in the interior of $R$ if the point representing it has $6$ neighbors at distance $1/N$ in the mesh and all of them represent states in $R$. Otherwise we will say that it is a border point of $R$.

The following result is a generalization of a familiar property of the solutions of the Laplace equation. It will be used in proving uniqueness of the solutions for the fixation probabilities:
\begin{proposition}[Maximum and minimum property]\label{propo 1}
	Let $R \subset \Lambda_N$ be connected. Then the maximum and the minimum of the fixation probability function $\alpha$ restricted to $R$ are on the border of $R$. The same property holds also for the maximum and the minimum of $\beta$ and $\gamma$.
\end{proposition}

\textbf{Proof}
	Assume that the fixation probability function $\alpha$, reaches its maximum at a state $(i_{\ast},j_{\ast})$ in the interior of $R$.
	
	We know that $\alpha_{i_{\ast}j_{\ast}}$ is the weighted average of the function $\alpha$ in the $6$ nearest neighbors of $(i_{\ast},j_{\ast})$ on $\Lambda_N$, all of which are in $R$. Being an average, $\alpha_{i_{\ast}j_{\ast}}$ cannot be strictly larger than any of the values of $\alpha$ at all the 6 nearest neighbors. As it is the maximum in the set with elements $(i_{\ast},j_{\ast})$ and its 6 nearest neighbors, then the values of $\alpha$ at all these 7 points must be the same. If we repeat the argument for the nearest neighbors of the nearest neighbors, and so on, we conclude that the occurrence of a maximum of $\alpha$ at an interior point implies that $\alpha$ is constant on $R$, so that the maximum also occurs at the border.
	
	A similar argument holds for the minimum of $\alpha$ and also for $\beta$ and $\gamma$.$\blacksquare$

We can now prove
\begin{theorem}[Uniqueness of $\alpha$, $\beta$ and $\gamma$]
	The linear system of equations (\ref{0}) for $(i,j) \in \Lambda_N$ with boundary conditions (\ref{0boundcond}), (\ref{1}) and (\ref{2}) has a unique solution. The same holds also for the analogous systems for $\beta$ and $\gamma$.
\end{theorem}

\textbf{Proof}
	If we order in some way the points of $\Lambda_N$, then the linear system for $\alpha$ may be written in matrix form $EX=F$, where $E$ is a square matrix of dimension $(N-1)(N-2)/2$ and $F$ is a column matrix depending only on the boundary conditions (\ref{0boundcond}), (\ref{1}) and (\ref{2}). 
	
	We claim that the only solution of the corresponding homogeneous linear system $EX = 0$ is the trivial $X=0$. In fact, the set of solutions for this homogeneous system is non-empty and the maximum and minimum property holds for these solutions. As the boundary condition for $EX=0$ is 0 at all border points of $\Lambda_N$, then the maximum and minimum of the solutions of $EX=0$ must be 0. Thus, the only solution is the trivial one.
	
	As a consequence, we must have $\det E \neq 0$. It follows that $E$ is invertible and $EX=F$ has a unique solution.
	
	Uniqueness of $\beta$ and $\gamma$ follow because the same matrix $E$ appears as the coefficient matrix of the corresponding linear systems. $\blacksquare$

\section{Coupling results}
\label{seccoupling}
We have already seen that the fixation probabilities for the Moran process with three strategies can be calculated by solving a linear system, but we no longer have an explicit formula for the solution, as (\ref{piiformula}) in the case of two strategies. We will show however that we can provide upper and lower bounds for these probabilities. These bounds will be derived by the \textit{coupling method} \cite{hollander}. 

Coupling  is a powerful way of comparing  two or more random variables by constructing them simultaneously through the same random device. More specifically, for a given Moran process with three strategies, which we call \textit{target chain}, we will construct a birth-death process as defined in Sect. \ref{secMorandef}, which we call \textit{comparison chain}, in which the fixation probability is explicitly calculated. We may realize the two chains simultaneously, and, as will be seen, this will give us bounds for the fixation probability in the target chain in terms of the exactly calculated fixation probability in the comparison chain.

In what follows, we will use the notations previously introduced for the transition probabilities both for birth-death processes and Moran processes with three strategies. More concretely, $a_i$ and $b_i$ will stand respectively for birth and death probabilities at state $i$ in a birth-death process, and $p^{XY}_{ij}$ are transition probabilities in the Moran process with three strategies introduced in (\ref{pAB}) and further equations. We also introduce
\begin{equation}
\label{defzij}
Z_{ij}^{+} \,=\, p^{AB}_{ij} +  p^{AC}_{ij} \;\;\; \textrm{and}  \;\;\; Z_{ij}^{-} \,=\, p^{BA}_{ij} +  p^{CA}_{ij} 
\end{equation}
for the probabilities of respectively increasing and decreasing the number of A individuals when the state is $(i,j)$.

We start with a general result:
\begin{theorem}  \label{lowboundtheo}
	Consider a Moran process with three strategies.	Suppose that there exists a birth-death process with states $\{0,1,2, \dots, N\}$ such that for all $(i,j) \in \Lambda_N$ the birth and death probabilities satisfy 
	\begin{equation}  \label{birthdeathcond1}
	a_i \leq Z_{ij}^{+} \;\;\;\textrm{and}  \;\;\; b_i \geq Z_{ij}^{-}
	\end{equation}
	and also
	\begin{equation}  \label{i-1cond}
	a_{i-1} \leq 1-Z_{ij}^{-}\;.
	\end{equation}
	If $\alpha_{ij}$ denotes the fixation probability of the strategy A and $\pi_i$ is the fixation probability at state $N$ in the birth-death process, then 
	\begin{equation}	\label{lowbound}
	\alpha_{ij} \geq \pi_i
	\end{equation} 
	for all $(i,j) \in \Lambda_N$.
\end{theorem}

In the above result, the target chain is the Moran process with three strategies and the comparison chain is the birth-death process. We may calculate the fixation probability $\pi_i$ of the comparison chain by  (\ref{piiformula}), thus obtaining a lower bound for the fixation probability $\alpha_{ij}$ of the Moran process with three strategies.

\textbf{Proof}
For each $(i,j) \in \Lambda_N$, consider the following mutually disjoint intervals, all contained in $[0,1]$: $I_1(i,j)=[0,q_1(i,j))$, $I_k(i,j)=[q_{k-1}(i,j),q_k(i,j))$, $k=2, 3, \dots, 6$ and $I_7(i,j)=[q_6(i,j),1]$, where $q_1(i,j)= p_{ij}^{AB}$, $q_2(i,j)=q_1(i,j)+p_{ij}^{AC}$, $q_3(i,j)=q_2(i,j)+p_{ij}^{BC}$, $q_4(i,j)=q_3(i,j)+ p_{ij}^{const}$, $q_5(i,j)=q_4(i,j)+p_{ij}^{CB}$, $q_6(i,j)=q_5(i,j)+p_{ij}^{CA}=1-p_{ij}^{BA}$. Notice that $\bigcup_{k=1}^7 I_k(i,j)=[0,1]$, that the sum of the lengths of $I_1(i,j)$ and $I_2(i,j)$ is $Z_{ij}^{+}$ and that the sum of the lengths of $I_6(i,j)$ and $I_7(i,j)$ is equal to $Z_{ij}^{-}$.

For each $i \in S$, we construct another set of mutually disjoint intervals contained in $[0,1]$ related to the comparison chain: $J_1(i)=[0,a_i)$, $J_2(i)=[a_i,1-b_i)$ and $J_3(i)=[1-b_i,1]$.
Conditions (\ref{birthdeathcond1}) imply 
\begin{equation} \label{ijinclusions}
J_1(i) \subset I_1(i,j) \cup I_2(i,j) \;\;\;\textrm{and}\;\;\; J_3(i) \supset I_6(i,j) \cup I_7(i,j)\;.
\end{equation}

The following vectors are the possible displacements in the state of the target chain: $d_1=(1,-1)$, $d_2=(1,0)$, $d_3=(0,1)$, $d_4= (0,0)$, $d_5=(0,-1)$, $d_6=(-1,0)$ and $d_7=(-1,1)$.

Suppose that at time 0 the states of the target and comparison chains are respectively $X_0=(i_0,j_0)$ and $Y_0=i_0$. The coupling of the target and comparison chains is accomplished by a sequence of independent uniformly distributed random variables $U_1, U_2, U_3, \dots \in [0,1]$ which will determine the state of both chains at all subsequent times. 

The state of both chains at time 1 will be obtained by displacements calculated as functions of $U_1$, then at time 2 by displacements calculated as functions of $U_2$, and so on. The way these displacements are calculated is as follows.

We declare that if the state of the target chain at time $\ell-1$ is $(i_{\ell-1},j_{\ell-1})$, then the $\ell$-th displacement of the target chain will be $d_k$ if $U_{\ell} \in I_k(i_{\ell-1},j_{\ell-1})$, $k=1,2, \dots, 7$, $\ell=1,2, \dots$. For the comparison chain, we declare that if its state at time $\ell-1$ is $i'_{\ell-1}$, $\ell=1,2, \dots$, then the displacement of the state of the comparison chain will be 1, 0, or -1, respectively, if $U_{\ell}$ is in $J_1(i'_{\ell-1})$, $J_2(i'_{\ell-1})$ or $J_3(i'_{\ell-1})$. Notice that the construction up to now is such that the probabilities of the possible displacements of both chains are all correctly distributed according to the chains' transition probabilities.

A fundamental observation is that, due to (\ref{ijinclusions}), whenever $U_1$ is such that there is a birth in the comparison chain, then the number of A individuals in the target chain will increase. And also, whenever $U_1$ is such that the number of A individuals decreases in the target chain, then there is a death in the comparison chain. As $i'_0=i_0$, it follows that $i_1 \geq i'_1$. 

We will prove by induction that $i_k \geq i'_k \; \forall k \in \mathbb{N}$. Suppose that $i_k \geq i'_k$ for a certain $k \in \mathbb{N}$. By the same reasoning used in proving that $i_1 \geq i'_1$, we see that if $i_k = i'_k$, then $i_{k+1} \geq i'_{k+1}$. If $i_{k} \geq i'_k+2$, then the conclusion $i_{k+1} \geq i'_{k+1}$ also holds, because $i_{k+1} \geq i_k-1$ and $i'_{k+1} \leq i'_k+1$. The only case in which it remains to prove that $i_{k+1} \geq i'_{k+1}$ is when $i_{k} = i'_k+1$. In this case, we use condition (\ref{i-1cond}), which we had not used, yet. This condition proves that if $U_k$ is such that a birth occurs in the comparison chain, then the number of A individuals in the target chain will not decrease and we will still have $i_{k+1} \geq i'_{k+1}$.

We have thus realized simultaneously the target and comparison chains according to their respective transition matrices in a way such that the initial states are respectively $(i_0,j_0)$ and $i_0$ and whenever there is fixation at state $N$ for the comparison chain, then there will be fixation of strategy A in the target chain. Thus $\alpha_{i_0,j_0} \geq \pi_{i_0}$. As $i_0$ is arbitrary, the theorem is proved. $\blacksquare$

An analogous result can be used to find an upper bound for the fixation probability in a Moran process with three strategies.
\begin{theorem}  \label{upperboundtheo}
	Consider a Moran process with three strategies and the same notation introduced before Theorem \ref{lowboundtheo}. If there exists a birth-death process with states $\{0,1,2, \dots, N\}$ such that for all $(i,j) \in \Lambda_N$ the birth and death probabilities  satisfy 
	\begin{equation}  \label{birthdeathcond2}
	a_i \geq Z_{ij}^{+} \;\;\;\textrm{and}  \;\;\; b_i \leq Z_{ij}^{-}
	\end{equation}
	and also
	\begin{equation}  \label{i+1cond}
	b_{i+1} \leq 1-Z_{ij}^{+}\;,
	\end{equation}
 then 
	\begin{equation}	\label{upperbound}
	\alpha_{i,j} \leq \pi_i
	\end{equation} 
	for all $(i,j) \in \Lambda_N$.
\end{theorem}

At this point the reader may wonder if we can in fact find comparison chains satisfying the hypotheses in Theorems \ref{lowboundtheo} and \ref{upperboundtheo}, so that upper or lower bounds for Moran processes with three strategies are produced. We show now that such comparison chains do exist in the important case of a Moran process with three strategies and frequency independent fitnesses. Although frequency independent fitnesses may be thought of as too much trivial, we know no other bounds for this particular case. Moreover, it will be seen that the following result will suggest how to obtain comparison chains for the general case.

\begin{theorem}\label{theofreqindep}
	Consider a Moran process for three types of individuals A, B and C and population size $N$, with frequency-independent fitnesses respectively given by $f>0$, $g>0$ and $h>0$. Suppose without loss of generality that $f>g>h$. Then
	the following bounds hold for all $(i,j) \in \Lambda_N$:
	\begin{eqnarray}
	\frac{1-\left(\frac{f}{g}\right)^{-i}}{1-\left(\frac{f}{g}\right)^{-N}} &\leq \alpha_{ij}& \leq \frac{1-\left(\frac{f}{h}\right)^{-i}}{1-\left(\frac{f}{h}\right)^{-N}} \label{alphabounds} \\
	\frac{1-\left(\frac{g}{f}\right)^{-j}}{1-\left(\frac{g}{f}\right)^{-N}} &\leq \beta_{ij}& \leq \frac{1-\left(\frac{g}{h}\right)^{-j}}{1-\left(\frac{g}{h}\right)^{-N}} \label{betabounds} \\
	\frac{1-\left(\frac{h}{f}\right)^{-(N-i-j)}}{1-\left(\frac{h}{f}\right)^{-N}} &\leq \gamma_{ij}& \leq \frac{1-\left(\frac{h}{g}\right)^{-(N-i-j)}}{1-\left(\frac{h}{g}\right)^{-N}}\;. \label{gammabounds}
	\end{eqnarray}	
\end{theorem}

Inequalities $f>g>h$ state that A individuals are fitter than Bs, which in turn are fitter than Cs. Focusing now on A individuals, the intuition behind the bounds above is that it becomes easier for As to fixate if we replace all Bs by Cs. The upper bound for $\alpha_{ij}$ in (\ref{alphabounds}) is just the fixation probability for A in a population with $i$ A individuals and $N-i$ Cs calculated by (\ref{freqindeppi}). Similarly, it is harder for As to fixate if we replace Cs by Bs, and the lower bound in (\ref{alphabounds}) is just the fixation probability for a population of $i$ As and $N-i$ Bs. The upper and lower bounds in (\ref{betabounds}) and (\ref{gammabounds}) are analogous. The rigorous proof for this intuition uses Theorems \ref{lowboundtheo} and \ref{upperboundtheo} with comparison chains obtained replacing individuals of one type by individuals of the other two types. In order to prove Theorem \ref{theofreqindep} and other results ahead, we will need the following result:
\begin{proposition}  \label{propminmaxz}
	Let $Z_{ij}^{\pm}$ be defined as in (\ref{defzij}). Then, for each fixed value of $i$, the minimum of $Z_{ij}^+$ and the maximum of $Z_{ij}^-$ for $j \in \{0,1,\dots, N-i\}$ are attained at the same value of $j$. Also, the maximum of $Z_{ij}^+$ and the minimum of $Z_{ij}^-$ for $j \in \{0,1,\dots, N-i\}$ are attained at the same value of $j$.
\end{proposition}	

\textbf{Proof}
	Just notice that $ Z_{ij}^{+} = i\frac{f_{ij}}{S_{ij}}\frac{N-i}{N}$ and $ Z_{ij}^{-} = \frac{j g_{ij}+(N-i-j)h_{ij}}{S_{ij}}\frac{i}{N}$ may be rewritten as  $(1-i\frac{f_{ij}}{S_{ij}})\frac{i}{N}$. For fixed $i$ the value of $j$ minimizing $i \frac{f_{ij}}{S_{ij}}$ will maximize $1-i\frac{f_{ij}}{S_{ij}}$.$\blacksquare$

We can now finally prove Theorem \ref{theofreqindep}:

\textbf{Proof}
Let $a_i$ be the probability of increasing the number of A individuals from $i$ to $i+1$ in a population with only A and B individuals. We also define $b_i$ as the probability of decreasing the number of A individuals from $i$ to $i-1$ in a population with only A and B individuals. As in (\ref{defzij}), let $Z_{ij}^{+}$ and $Z_{ij}^{-}$ be respectively the probabilities of increasing and decreasing the number of A individuals from $i$ to $i\pm 1$ in a population with A, B and C individuals and frequency independent fitnesses. We have
\begin{align}
a_{i} & = \frac{if}{if + (N-i)g}\frac{N-i}{N}\;, & Z_{ij}^{+} & = \frac{if}{if + jg + (N-i-j)h}\frac{N-i}{N}\;, \nonumber \\
b_{i} & = \frac{(N-i)g}{if + (N-i)g}\frac{i}{N}\;, & Z_{ij}^{-} & = \frac{jg + (N-i-j)h}{if + jg + (N-i-j)h}\frac{i}{N}\;.\nonumber
\end{align}

As $g>h$, then $if + (N-i)g= if + jg + (N-i-j)g \geq if + jg + (N-i-j)h$ for all $(i,j)\in \Lambda_N$. It follows that 
\begin{equation}   \label{jmin}
a_i= Z_{i,N-i}^{+} \leq Z_{ij}^{+}
\end{equation} 
for all $(i,j)\in \Lambda_N$.

By Proposition \ref{propminmaxz}, $b_i= Z_{i,N-i}^{-}\geq Z_{ij}^{-}$ for all $(i,j)\in \Lambda_N$. We have thus proved that conditions (\ref{birthdeathcond1}) in Theorem \ref{lowboundtheo} are fulfilled. The lower bound in (\ref{alphabounds}) will result if we prove that (\ref{i-1cond}) is true.

In fact, it can be seen, after some tedious manipulations, that
\begin{eqnarray*}
	&& 1 - Z_{ij}^{-} - a_{i-1} \;=\; \\ 
	& = & \frac{1}{N S_{ij}S_{i-1,N-i+1}} \, \left\{ i(i-1)^{2}f^{2} + (N-i+1)(N-i)g \,S_{ij} \right.\\ 
	&+& \left. if[(N-i-j)(g-h) + [(i-1)(N-i)+i]g]\,+ [jg+(N-i-j)h]f\right\}\;. 
\end{eqnarray*}
As all terms around the curly brackets in the above expression are obviously  non-negative, as well as the denominator $N S_{ij}S_{i-1,N-i+1}$, then condition (\ref{i-1cond}) is satisfied and the lower bound in (\ref{alphabounds}) proved.

All the remaining bounds can be proved in an analogous way, either using Theorem \ref{lowboundtheo} or Theorem \ref{upperboundtheo}.
$\blacksquare$

An interesting consequence of Theorem \ref{theofreqindep} concerns the behavior of the fixation probabilities $\alpha$, $\beta$ and $\gamma$ when population size $N$ tends to infinity. For large populations we expect that the randomness inherent in the Moran process becomes less important and, if the Moran process is compatible with natural selection, only the fittest individuals should survive. In order to prove this compatibility, it is necessary that the fractions $x$, $y$ and $1-x-y$ of A, B and C individuals are fixed, whereas the respective numbers thereof tend to infinity. This idea is precisely defined if we define for
$(x,y) \in \Lambda$
\begin{eqnarray}
A_{N}(x,y) & = & \alpha_{[Nx],[Ny]} \;,\nonumber \\
B_{N}(x,y) & = & \beta_{[Nx],[Ny]} \label{defprobfrac}\;, \\
\Gamma_{N}(x,y) & = & \gamma_{[Nx],[Ny]} \nonumber\;,
\end{eqnarray}
where $[z]$ denotes the integer closest to $z$ and the set $\Lambda$ was defined in (\ref{defLambda}).

\begin{corollary}\label{corofreqindep}
Consider the same hypotheses of Theorem \ref{theofreqindep}. Then, for any $(x,y) \in \Lambda$,  $A_{N}(x,y)\stackrel{N \rightarrow \infty}{\rightarrow} 1$, $B_{N}(x,y)\stackrel{N \rightarrow \infty}{\rightarrow} 0$ and $\Gamma_{N}(x,y)\stackrel{N \rightarrow \infty}{\rightarrow} 0$.
\end{corollary}

\textbf{Proof}
By  Theorem \ref{theofreqindep}, $$A_{N}(x,y) \geq \frac{1-\left(\frac{g}{f}\right)^{[Nx]}}{1-\left(\frac{g}{f}\right)^{N}}\;.$$ 
The result for $A_N$ is proven by taking the limit $N \rightarrow \infty$ and using $f > g>0$. The results for $B_N$ and $\Gamma_N$ follow because they are both probabilities and their sum with $A_N$ equals 1.
$\blacksquare$

The following theorem is a generalization of the ideas presented in Theorem \ref{theofreqindep} and works as a general ``recipe'' for  constructing comparison chains satisfying hypotheses of Theorems \ref{lowboundtheo} and \ref{upperboundtheo}.  Contrarily to Theorem \ref{theofreqindep}, the proof for the next result requires that the population size $N$ is large enough. We are not sure whether this is a necessary condition.

\begin{theorem}\label{theogeneralbounds}
Consider a Moran process with three strategies and the notation introduced in Theorem \ref{lowboundtheo}. Define a comparison birth-death process with birth and death probabilities respectively given by
\begin{equation}\label{lowbounddef}
a_{i}^{low}\,=\, \min_{0 \leq j \leq N-i} Z_{ij}^{+} \;\;\;\mathrm{and}\;\;\;
b_{i}^{low} \,=\, \max_{0 \leq j \leq N-i} Z_{ij}^{-}\;.
\end{equation}
Let $\pi_{i}^{low}$ denote the fixation probability in state $N$ of the comparison birth-death process and $\alpha_{ij}$ denote the Moran process fixation probability for type A in the initial state $(i,j)$. Then, for large enough $N$,
\[\alpha_{ij} \geq \pi_i^{low}\;.\]

Similarly, if we define another birth-death with fixation probability $\pi_i^{up}$ by taking
\begin{equation}\label{upperbounddef}
a_{i}^{up} \,=\, \max_{0 \leq j \leq N-i} Z_{ij}^{+} \;\;\;\mathrm{and}\;\;\;
b_{i}^{up}\,=\, \min_{0 \leq j \leq N-i} Z_{ij}^{-} \;,
\end{equation}
then, for large enough $N$, \[\alpha_{ij} \leq \pi_i^{up}\;.\]
\end{theorem}

\textbf{Proof}
We will show that the birth-death process (\ref{lowbounddef}) satisfies the hypotheses of Theorem \ref{lowboundtheo} for $N$ large enough. The proof that the process defined by (\ref{upperbounddef}) satisfies the hypotheses of Theorem \ref{upperboundtheo} for large enough $N$ is analogous.

By (\ref{lowbounddef}), we automatically have for each $i$ that $a_{i}^{low} \leq Z_{ij}^{+}$ and $b_{i}^{low} \geq Z_{ij}^{-}$ for all $j$ such that $(i,j) \in \Lambda_N$. To complete the proof, we need to show that $a_{i-1}^{low} \leq 1 - Z_{ij}^{-}$ if $N$ is large enough.

To see that, we write
\begin{equation}  \label{1-z-z+}
1-Z_{ij}^--Z_{i-1,j}^+ \,=\,(1-Z_{ij}^--Z_{ij}^+) + (Z_{ij}^+ - Z_{i-1,j}^+)\;.
\end{equation} 
The first term $1-Z_{ij}^--Z_{ij}^+$ is the probability at state $(i,j)$ that the number of A individuals remains constant. It can be written as the sum $p_{ij}^{AA}+(p_{ij}^{BC}+p_{ij}^{CB}+p_{ij}^{BB}+p_{ij}^{CC})$, in which $p_{ij}^{AA}$ vanishes only if $i=0$ and the sum of the remaining four terms vanishes only if $i=N$. 

Writing $x=i/N$ and $y=j/N$ and using a reasoning similar to the one exemplified in (\ref{fijlimit}), we get
\begin{equation}  \label{asymp1-z-z}
1-Z_{ij}^--Z_{ij}^+ \,=\, C_1(x,y)+C_2(x,y) + O(\frac{1}{N}) \;,
\end{equation}
where
\[C_1(x,y) \,=\,
\frac{x^2 F(x,y)}{xF(x,y)+y G(x,y)+(1-x-y) H(x,y)} \]
comes from $p_{ij}^{AA}$ and
\[C_2(x,y) \,=\, \frac{(1-x) (yG(x,y)+ (1-x-y) H(x,y)) }{xF(x,y)+y G(x,y)+(1-x-y) H(x,y)} \]
comes from the sum $p_{ij}^{BC}+p_{ij}^{CB}+p_{ij}^{BB}+p_{ij}^{CC}$. 

Observe that both $C_1$ and $C_2$ are continuous functions with values in $[0,1]$ in the compact triangle $\Lambda$ defined in (\ref{defLambda}). Moreover $C_1(1,0)=1$ and $C_2(0,y)=1$, so that there exist $x_1,x_2 \in[0,1]$, $x_1<x_2$ such that $C_2(x,y) \geq 1/2$ if $(x,y)\in \Lambda$ with $x \leq x_1$ and $C_1(x,y) \geq 1/2$ if $(x,y)\in \Lambda$ with $x \geq x_2$. As neither $C_1$ and $C_2$ vanishes for the points $(x,y) \in \Lambda$ with $x \in[x_1,x_2]$, then their sum has a positive minimum  value $\mu$ in this set. Of course the minimum value of $C_1+C_2$ in $\Lambda$ cannot be smaller than the smallest between $\mu$ and $1/2$, being then positive and independent of $N$. This proves that $1-Z_{ij}^--Z_{ij}^+$ is bounded away from 0 for large enough $N$.

Using the same ideas,
\[Z_{ij}^+\,=\, D(x,y)+O(\frac{1}{N})\;, \]
with
\[D(x,y) \,=\, \frac{x(1-x) F(x,y)}{xF(x,y)+y G(x,y)+(1-x-y) H(x,y)}\;.\]
Then the second summand in the right-hand side of (\ref{1-z-z+}) becomes
\begin{eqnarray}
Z_{ij}^+ - Z_{i-1,j}^+ &=& D(x,y)-D(x-\frac{1}{N},y) + O(\frac{1}{N})\nonumber \\
&=& -\frac{1}{N} \frac{\partial D}{\partial x}(x,y) +O(\frac{1}{N})\;.\nonumber
\end{eqnarray}

We have thus shown that one of the terms in the right-hand side of (\ref{1-z-z+}) is positive and $O(1)$, and the other is $O(\frac{1}{N})$. This proves that $1-Z_{ij}^--Z_{i-1,j}^+>0$ for all $(i,j)\in \Lambda_N$ for large enough $N$ and the proof is completed. 
$\blacksquare$

\section{Strict Nash equilibria and related results}
\label{secnash}
In general, consequences of Theorem \ref{theogeneralbounds} depend on knowing for each $i$ the location of the maximum or of the minimum among the values $Z^+_{ij}$, $j=1, 2, \dots, N-i$. As we will see in a concrete example in Sect. \ref{seccoop}, this may be a complicated task. The results in this section refer to important situations in which Theorem \ref{theogeneralbounds} may be used without the need of locating the maximum or the minimum of the $Z^+_{ij}$.

One situation is the $N \rightarrow \infty$ limit of the fixation probability of a strategy when this strategy is a strict Nash equilibrium and its population frequency is close to 1. We remind that strategy A is a \textit{strict Nash equilibrium} (see e.g. \cite{hofbauersigmund} or \cite{nowakbook}) if $m_{11}> m_{i1}$ for $i=2$ and $i=3$. It can be shown that if A is a strict Nash equilibrium, then the point $(1,0) \in \Lambda$ corresponding to the whole population being of type A is an \textit{asymptotically stable} equilibrium for the replicator dynamics. In other words, every orbit of the replicator dynamics which starts close enough to point $(1,0)$ will end in that point. As an important similarity of the Moran process with deterministic dynamics we will show that if strategy A is a strict Nash equilibrium, then there exists a neighborhood of $(1,0)$ in $\Lambda$ such $A_N(x,y) \stackrel{N \rightarrow \infty}{\rightarrow} 1$ for $(x,y)$ in this neighborhood. 

As we will shortly see, the above claim will follow as a consequence of this more general result:
\begin{theorem}\label{theobeforeAstrictNash}
Consider a Moran process with three strategies. Suppose there exist $s>1$, $N^* \in \mathbb{N}$ and $x^* \in[0,1)$ such that if $N \geq N^*$ and $\frac{i}{N}>x^*$, then
\[\frac{Z^+_{ij}}{Z^-_{ij}} \geq s\]
holds $\forall j \in \{0, 1,\dots, N-i\}$. Then 
\[\lim_{N \rightarrow \infty}A_N(x,y) =1\] 
for all $(x,y) \in \Lambda$ with $x>x^*$.
\end{theorem}

\textbf{Proof}
Let $x \in (0,1)$ and consider the lower bound comparison birth-death process defined in Theorem \ref{theogeneralbounds} by (\ref{lowbounddef}). By Proposition \ref{propminmaxz}, we know that the maximum over $j$ of $Z_{[Nx],j}^{-}$ and the minimum over $j$ of $Z_{[Nx],j}^{+}$ occur at the same value $\overline j(x) \in \{0, 1, \dots, [Nx]\}$. In other words,
\[r_{[Nx]}^{low} \equiv \frac{a_{[Nx]}^{low}}{{b_{[Nx]}^{low}}}= \frac{Z^{+}_{[Nx],\overline j(x)}}{Z^{-}_{[Nx],\overline j(x)}}.\]  

Suppose now that $x>x^*$ and take $N \geq N^*$ and also large enough so that $[Nx]/N>x^{*}$. Then $r^{low}_{[Nx]}$ is strictly greater than $s$ for all $x>x^{*}$. By Theorem \ref{partialfixationtheo} in Appendix \ref{appbdprocesses}, we conclude that $\lim_{N \rightarrow \infty}\pi^{low}_{[Nx]}=1$ for all $x>x^{*}$. As, by Theorem \ref{theogeneralbounds}, $\pi^{low}_{[Nx]}\leq \alpha_{[Nx],[Ny]}$, the theorem is proved. 
$\blacksquare$

We can now prove our important result concerning the case of strategy A being a strict Nash equilibrium:
\begin{theorem}\label{theoAstrictNash}
Consider a Moran process with three strategies such that strategy A is a strict Nash equilibrium. Then there exists $x^* \in[0,1)$ such that $\lim_{N \rightarrow \infty}A_N(x,y) = 1$ for all $(x,y) \in \Lambda$ with $x>x^{*}$.
\end{theorem}

\textbf{Proof}
We will show that there exist $x^*$, $s$ and $N^*$ as in the hypotheses of Theorem \ref{theobeforeAstrictNash}. The result will then follow as a consequence of that theorem.

In fact, if strategy A is a strict Nash equilibrium, then $F(1,0)>G(1,0)$ and $F(1,0)>H(1,0)$, see (\ref{detfitnesses}), and, by continuity, we have a neighborhood of $(1,0)$ in $\Lambda$ in which the deterministic fitness $F$ is strictly larger than both $G$ and $H$. 
	
Let $x_1$ be the greatest lower bound of the values $x \in [0,1]$ such that $F(x,y)>G(x,y)$  and $F(x,y)>H(x,y)$  hold simultaneously for all $y$ such that $(x,y) \in \Lambda$.

In analogy with what we did in (\ref{asymp1-z-z}), we may rewrite $Z^{+}_{ij}/Z^{-}_{ij}$ as an asymptotic term $R(x,y)$, where $x=i/N$ and $y=j/N$, plus corrections that tend to 0 as $N \rightarrow \infty$. We obtain
\begin{equation}  \label{defR}
R(x,y) = \frac{(1-x)F(x,y)}{yG(x,y)+(1-x-y)H(x,y)}\;,
\end{equation}
which is continuous in $\Lambda \setminus (1,0)$. 

Choose $x^*\in (x_1,1)$ and define $\Lambda^* = \{(x,y) \in \Lambda\, ; x^* \leq x < 1\}$. If we define $R^* = \inf_{(x,y) \in \Lambda^*}R(x,y)$, we claim that $R^*>1$.

In fact, although $R$ is not defined at $(1,0)$, both $F/G$ and $F/H$ are continuous at this point. So, we define $S(x,y) = \min\left\{\frac{F(x,y)}{G(x,y)},\frac{F(x,y)}{H(x,y)}\right\}$, which is continuous in the compact set $\overline{\Lambda^*} =\Lambda^* \cup \{(1,0)\}$. Let $s^*$ be the minimum value of $S$ on $\overline{\Lambda^*}$. As $F(x,y)>G(x,y)$ and $F(x,y)>H(x,y)$ in $\overline{\Lambda^*}$, then $s^*>1$. Moreover, $R(x,y) \geq S(x,y)$ for $(x,y) \in \Lambda^*$. Thus $R^* \geq s^*$, proving our claim that $R^*>1$.

We will now estimate the difference  between $Z^{+}_{ij}/Z^{-}_{ij}$ and $R(\frac{i}{N}, \frac{j}{N})$. Using (\ref{defzij}), we have
\[\frac{Z^+_{ij}}{Z^-_{ij}} \,=\, \frac{(1-\frac{i}{N})f_{ij}}{\frac{j}{N} g_{ij}+ (1-\frac{i}{N}-\frac{j}{N})h_{ij}} \;.\]
Using also the definition (\ref{defR}) of $R$, we get 
\begin{align}
&\frac{Z^+_{ij}}{Z^-_{ij}}-R(\frac{i}{N}, \frac{j}{N}) =\\
&=\frac{(1-\frac{i}{N})\left\{f_{ij}[\frac{j}{N}G(\frac{i}{N},\frac{j}{N})+ (1-\frac{i}{N}-\frac{j}{N})H(\frac{i}{N},\frac{j}{N})] - F(\frac{i}{N},\frac{j}{N})[\frac{j}{N}g_{ij}+ (1-\frac{i}{N}-\frac{j}{N})h_{ij}]\right\}}{[\frac{j}{N} g_{ij}+ (1-\frac{i}{N}-\frac{j}{N})h_{ij}][\frac{j}{N}G(\frac{i}{N},\frac{j}{N})+(1-\frac{i}{N}-\frac{j}{N})H(\frac{i}{N},\frac{j}{N})]}\nonumber \\
&= \frac{(1-\frac{i}{N}) \, \phi_{i,j,N}}{[\frac{j}{N} g_{ij}+ (1-\frac{i}{N}-\frac{j}{N})h_{ij}][\frac{j}{N}G(\frac{i}{N},\frac{j}{N})+(1-\frac{i}{N}-\frac{j}{N})H(\frac{i}{N},\frac{j}{N})]}\;, \label{expphi}
\end{align}
where
\begin{align*}
\phi_{i,j,N} &=\,\left\{f_{ij}-F(\frac{i}{N},\frac{j}{N}) \right\}\left[\frac{j}{N}G(\frac{i}{N},\frac{j}{N})+ (1-\frac{i}{N}-\frac{j}{N})H(\frac{i}{N},\frac{j}{N})\right]+ \\
	&+ F(\frac{i}{N},\frac{j}{N}) \left\{\frac{j}{N}[G(\frac{i}{N},\frac{j}{N})-g_{ij}]+ (1-\frac{i}{N}-\frac{j}{N})[H(\frac{i}{N},\frac{j}{N})-h_{ij}]\right\}\;.
\end{align*}

If $M_1 \equiv \inf_{N \in \mathbb{N}}\min_{(i,j) \in \Lambda_N} \{g_{ij},h_{ij}, G(\frac{i}{N},\frac{j}{N}),H(\frac{i}{N},\frac{j}{N})\}$, then the denominator in (\ref{expphi}) is bounded below by $(1-\frac{i}{N})^2 M_1^2$. By the continuity in $\Lambda$ of $G$ and $H$ and by formulas analogous to (\ref{fijlimit}) for $G$ and $H$, we know that $M_1$ is finite and positive.

We can also find an upper bound for the  $\phi_{i,j,N}$ in the numerator. Let $M_2 \equiv \max_{(x,y) \in \Lambda} \{F(x,y), G(x,y),H(x,y)\}$. By (\ref{fijlimit}) and analogous expressions, there also exists a constant $c>0$ such that for all $N \in \mathbb{N}$,
\[\max_{(i,j) \in \Lambda_N}\{|F(\frac{i}{N},\frac{j}{N})-f_{ij}|,|G(\frac{i}{N},\frac{j}{N})-g_{ij}|,|H(\frac{i}{N},\frac{j}{N})-h_{ij}|\} \,<\, \frac{c}{N}\;.\]
Using these bounds, we get $|\phi_{i,j,N}|<(1-\frac{i}{N})\frac{2c M_2}{N}$.

Putting together the bounds for numerator and denominator in (\ref{expphi}), we can see that there exists a constant $K$ such that
\[\left|\frac{Z^+_{ij}}{Z^-_{ij}}-R(\frac{i}{N}, \frac{j}{N})\right| < \frac{K}{N}\;.\]

Let now $s=\frac{1}{2}(R^*+1)$ and $N^*$ be the smallest integer not smaller than $\frac{K}{R^*-s}$. Then, for $N>N^*$ and $\frac{i}{N}>x^*$ we have $\frac{Z^+_{ij}}{Z^-_{ij}} \geq s$ for all $j \in \{0,1, \dots, N-i\}$. By Theorem \ref{theobeforeAstrictNash} we conclude that $\lim_{N \rightarrow \infty}A_N(x,y) =1$ 
for all $(x,y) \in \Lambda$ with $x>x^*$. $\blacksquare$

Theorem \ref{probfix vertice teo zero} ahead deals with a situation which is in some sense inverse to that in Theorem \ref{theoAstrictNash}. We suppose there that $F(1,0)<G(1,0)$ and $F(1,0)<H(1,0)$. This, by continuity, implies that in a neighborhood of the point $(1,0)\in \Lambda$ strategy A is the least fit. In the replicator dynamics, this hypothesis implies that vertex A is a repeller of the dynamics. Despite that, it is not true that $A_N(x,y) \rightarrow 0$ as $N \rightarrow \infty$ if $x$ is close to 1.
 
Before enunciating Theorem \ref{probfix vertice teo zero}, we make a definition for Moran processes with three strategies analogous to another definition made in Appendix \ref{appbdprocesses} for birth-death processes. 

Let $i>i^*$. We define $\alpha_{i,j \backslash i^{*}}$ as the probability with initial condition $(i,j)$ that strategy A fixates without ever returning to any among the states $(i^{\ast},k)$ $k = 0, 1, ..., N-i^{*}$. In other words, in $\alpha_{i,j \backslash i^{*}}$ we take into account only events in which strategy A fixates and the number of A individuals is always larger than $i^*$. Similarly, for $x>x^*$ and $N$ large enough so that $[Nx]>[Nx^*]$, we define $A_{N \backslash x^{*}}(x,y) =\alpha_{[Nx],[Ny]\backslash [Nx^*]}$.

We start by stating a preparatory result analogous to Theorem \ref{theobeforeAstrictNash}:

\begin{theorem}\label{theobeforeArepeller}
	Consider a Moran process with three strategies. Suppose there exist $s \in (0,1)$, $N^* \in \mathbb{N}$ and $x^* \in[0,1)$ such that if $N \geq N^*$ and $\frac{i}{N}>x^*$, then
	\[\frac{Z^+_{ij}}{Z^-_{ij}} \leq s\]
	holds $\forall j \in \{0, 1,\dots, N-i\}$. Then 
	\[\lim_{N \rightarrow \infty}A_{N \backslash x^{*}}(x,y) =0\] 
	for all $(x,y) \in \Lambda$ with $x>x^*$.
\end{theorem}

We do not write a complete proof of Theorem \ref{theobeforeArepeller}, because it is analogous to the proof of Theorem \ref{theobeforeAstrictNash}, but we explain the important differences. First of all, instead of using a lower bound comparison birth-death process, we take an upper bound (\ref{upperbounddef}).  The conclusion is a consequence of Theorem \ref{partialfixationtheoanalogo} in Appendix \ref{appbdprocesses}. 

It is not possible to obtain the stronger result $A_N(x,y) \rightarrow 0$ as $N \rightarrow \infty$, because $A_N(x,y) \geq A_{N \backslash x^{*}}(x,y)$. Not only we are not able to prove that $A_N(x,y) \rightarrow 0$, but we can give an example in which the hypotheses of Theorem \ref{theobeforeArepeller} are fulfilled but we have for all $(x,y) \in \Lambda$ with $x>0$ that $A_N(x,y) \rightarrow 1$. Just take a pay-off matrix
\[M=\begin{pmatrix}
a&b&b\\c&d&d\\c&d&d
\end{pmatrix}\]
with $a, b, c$ and $d$ all positive, $a<c$ and $d<b$. The peculiar form of $M$ implies that individuals adopting strategies B and C have the same fitness for any population composition. As a result, for the sake of calculating the fixation probabilities it is as if we had only two strategies. Thus $\alpha_{ij}$ is independent of $j$ and can be calculated by (\ref{piiformula}). Inequality $a<c$ guarantees that $F(1,0)<G(1,0)=H(1,0)$, which, by continuity, implies $Z^+_{ij}/Z^-_{ij} \leq s<1$ for $i$ close to $N$. The other inequality $d<b$ implies that strategies B and C are not Nash equilibria. This is a hypothesis necessary for using Theorem 5 in \cite{graphshapes}. Using the notation of the above cited work, by taking $d$ close enough to 0 we get $L(1)<0$. According to Theorem 5 in \cite{graphshapes}, we will have $\lim_{N \rightarrow \infty} A_N(x,y) = 1$ for all $(x,y) \in \Lambda$ with $x>0$.

Due to the above example, the best result analogous to Theorem \ref{theobeforeAstrictNash} we can have is
\begin{theorem}\label{probfix vertice teo zero}
Consider a Moran process with three strategies. If $F(1,0) < G(1,0)$ and $F(1,0) < H(1,0)$, then there exists $x^{\ast} \in [0,1)$ such that 
\[\lim_{N \rightarrow \infty}A_{N \backslash x^{*}}(x,y) = 0\]
for all $(x,y) \in \Lambda$ with $x>x^{*}$.
\end{theorem}

We may omit the proof because it is just a repetition of the ideas in the proof of Theorem \ref{theoAstrictNash} of approximating $Z^+_{ij}/Z^-_{ij}$ by $R(\frac{i}{N}, \frac{j}{N})$ and then using Theorem \ref{theobeforeArepeller}.

The example after Theorem \ref{theobeforeArepeller} shows that if strategy A is the least fit and it is close to fixation, then for large $N$ its fixation probability may even be close to 1. This will not happen if A is the least fit strategy in a region away from its fixation. This is the content of our last result in this section.

\begin{theorem}\label{partialexttheo}
Consider a Moran process with three strategies. If there exists $x^{*}\in (0,1]$ such that $F(x,y) < G(x,y)$ and $F(x,y) < H(x,y) \ \forall \ (x,y) \in \Lambda$ with $x < x^{*}$, then $\lim_{N \rightarrow \infty}A_{N}(x,y) = 0$ for all $(x,y) \in \Lambda$ with $x<x^{*}$. 
\end{theorem}

\textbf{Proof}
We only sketch the proof, because it is again similar to preceding ones. We use the upper bound (\ref{upperbounddef}) for $\alpha_{ij}$ in Theorem \ref{theogeneralbounds}. In order to prove the thesis, we should show that for fixed $x<x^*$ we have $\pi^{up}_{[Nx]}\rightarrow 0$ when $N \rightarrow \infty$. Hypotheses $F(x,y) < G(x,y)$ and $F(x,y) < H(x,y)$ if $(x,y) \in \Lambda$ and $x<x^*$ make sure that $r_i^{up} \equiv \frac{a_i^{up}}{b_i^{up}}\leq s<1$ if $N$ is large enough and $i<Nx^*$.
	
The result that, for $x<x^*$,  $\pi^{up}_{[Nx]}\rightarrow 0$ when $N \rightarrow \infty$ may be attained in two equivalent ways. One is to develop for $i<i^*$ a result analogous to Proposition \ref{partialfixationprop} for the probability of fixation at state 0 of the comparison chain without attaining state $i^*$. The other way is proving by Theorem \ref{partialfixationtheo} that the fixation probability at state $N$ of the \textit{dual process}, see formulas (\ref{defdualprocess}) to (\ref{recorrenciapiibarra}), tends to 1 as $N \rightarrow \infty$.
$\blacksquare$

It is important to remember that, with a few exceptions, all the results presented since the beginning of Sect. \ref{seccoupling} refer to the fixation of the A strategy. By making appropriate adjustments in the hypotheses, analogous results are also valid for the fixation of strategies B and C.

\section{The evolution of cooperation with three strategies from a stochastic point of view}
\label{seccoop}

In \cite{nunezneves} some results were proved for the replicator dynamics in a model for the evolution of cooperation in a population with three strategies. In this section, after a brief description of the problem, we intend to use some of the results proved so far for the Moran process with three strategies to gain some understanding of the stochastic version of the results obtained in that paper.

\cite{nowaksigmundnature} considered a population with 100 types of individuals following different \textit{reactive strategies} for the infinitely repeated prisoner's dilemma (IRPD). In the prisoner's dilemma, individuals may at each interaction either cooperate or defect. Reactive strategies are characterized by two parameters: \textit{loyalty} and \textit{forgiveness}. The loyalty of a reactive strategy is the probability that the player adopting this strategy cooperates after receiving cooperation in the previous interaction. The forgiveness is the probability that the player cooperates after receiving a defection in the previous interaction. Although reactive strategies are characterized by probabilities, the pay-offs in the IRPD are deterministically calculated as an average of infinitely many interactions \cite{nowakbook}. Nowak and Sigmund numerically solved the replicator dynamics for this population, a system of 99 ordinary differential equations. Their numerical solution suggested that among the 100 strategies, only three play a prominent role in their numerical experiment. 

The first prominent strategy is ALLD: individuals which always defect. The second important strategy is ATFT (almost tit-for-tat): individuals with loyalty close to 1 and small positive forgiveness. The third strategy, which we will denote as G (generous), has loyalty equal to 1, and positive forgiveness $q$ larger than the forgiveness of the ATFT.

In the numerical experiment by \cite{nowaksigmundnature}, initially the population frequency of most strategies is strongly depleted, with the exception of strategies closest to ALLD. After this initial period, the frequency of strategies close to ATFT increased, almost attaining the whole population. But the ultimate winner in their simulations was a surprisingly cooperative strategy which they called GTFT (generous tit-for-that), i.e. a strategy of the G kind with an optimum value of $q$ which allows the followers of this strategy not to be too much exploited by defectors. This optimum value of $q$ turned out to be $1/3$ for the parameter values used in their experiment.

\cite{nunezneves} studied a simplified version of the extremely complicated population in \cite{nowaksigmundnature}. They considered only three kinds of individuals adopting the afore mentioned prominent reactive strategies: ALLD, ATFT and G. The forgiveness parameter $q$ of the G individuals may be varied. The results of \cite{nunezneves} show that, according to the value of $q$, there are several different scenarios for which strategies survive in the infinite time limit and, consequently, different types of evolution of cooperation, or non-evolution of cooperation, may occur. In particular, they prove that there exists a threshold value for $q$ under which the result of the numerical experiment by Nowak and Sigmund holds. More exactly, in a population with the three types of individuals above mentioned, existence is proved of a threshold $q_{GTFT}$ such that for $q<q_{GTFT}$ there will be a region with positive area such that for initial conditions in this region only G individuals will survive. When only G individuals survive the scenario is termed one in which \textit{full evolution of cooperation} holds.

\begin{figure}
	\begin{center}
		\includegraphics[width= 0.7\textwidth]{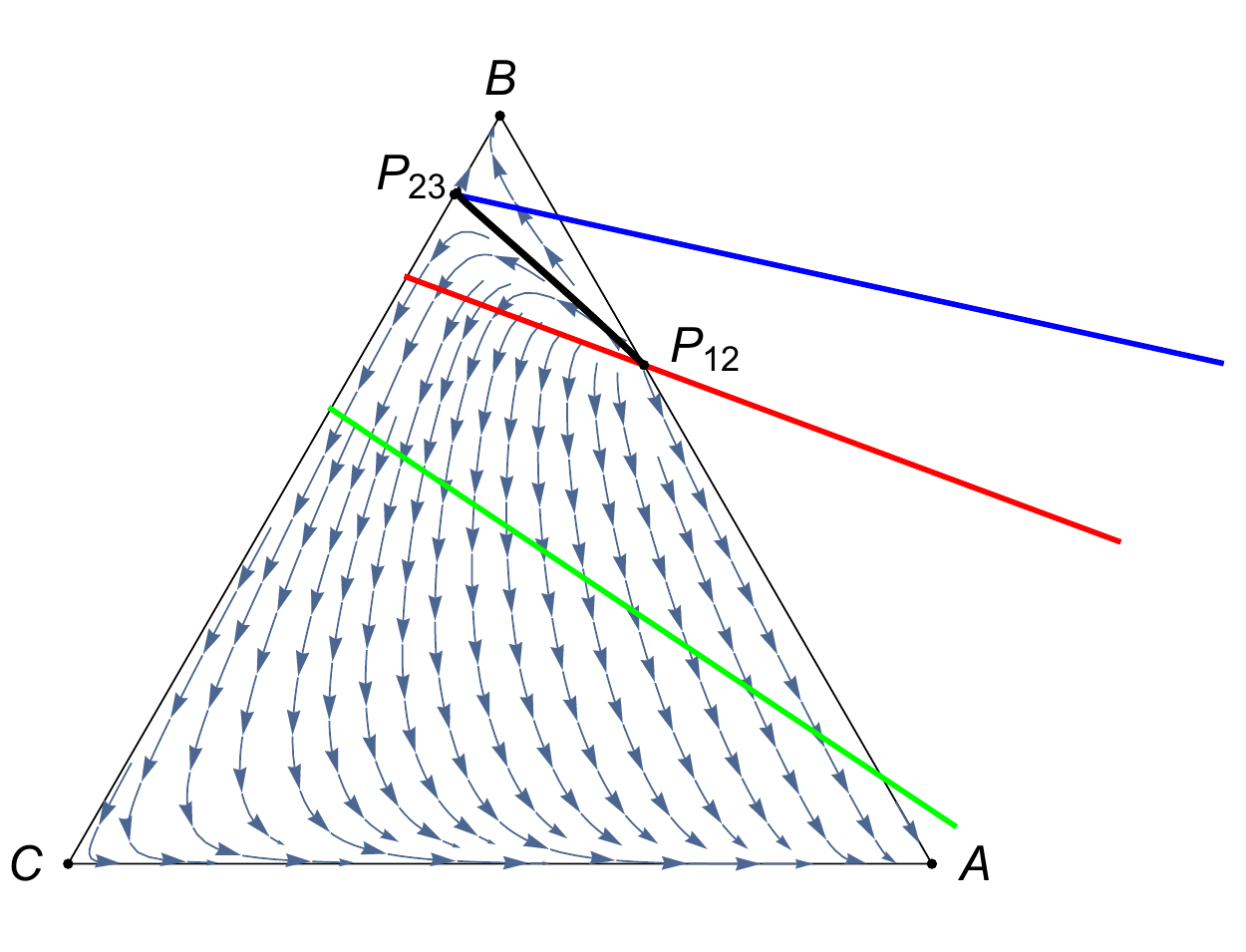}
		\caption{Some orbits for the replicator dynamics with pay-off matrix in (\ref{payofffullevo}). Strategy A is G (generous), strategy B is ALLD (always defect), and strategy C is ATFT (almost tit-for-tat). The parameters for the prisoner's dilemma (see \cite{nunezneves} or \cite{nowaksigmundnature}) are $T=5$, $R=3$, $P=1$, $S=0$. The G strategy has a forgiveness parameter $q=1/3$ and the ATFT has loyalty equal to 0.9 and forgiveness 0.1. Besides the vertices of the triangle, the only equilibria of the dynamics for this choice of parameters and strategies are the points $P_{12}$ and $P_{13}$. $P_{12}$ is a repeller and $P_{23}$ is a saddle point. All orbits inside the triangle and above the separatrix joining these two equilibria converge to vertex B and all orbits inside the triangle and below the same separatrix converge to vertex $A$. Strategies ALLD and G are strict Nash equilibria. Let $F$, $G$ and $H$ denote as elsewhere in this paper the deterministic fitnesses (\ref{detfitnesses}). The red line is the set of points in which  $F=G$. The green line is the set of points in which $F=H$. The blue line is the set of points in which $G=H$.
			\label{fullevodet}}
	\end{center}
\end{figure}

We will consider fixed numerical values for the parameters in \cite{nunezneves}, so that the pay-off matrix $M$ is also fixed. Our parameter choice is such that $q$ is below the threshold $q_{GTFT}$ for full evolution of cooperation. More concretely, the pay-off matrix considered is 
\begin{equation}  \label{payofffullevo}
	M = \left(
	\begin{array}{ccc}
		3.00 & 0.67 & 2.61 \\
		2.33 & 1.00 & 1.40 \\
		2.97 & 0.90 & 2.25 \\
	\end{array}
	\right) \;,
\end{equation}
where strategy A (numbered 1 in the matrix) is G, strategy B (numbered 2) is ALLD, and strategy C (numbered 3) is ATFT. The reader consulting \cite{nunezneves} should be aware that the numbering of strategies here differs with respect to that paper. Fig. \ref{fullevodet} shows some orbits of the replicator dynamics for the pay-off matrix (\ref{payofffullevo}). Complete specification of parameter choices and other useful information may be found in the caption of that figure.

Just by looking at the above pay-off matrix, we know that strategies A and B are both strict Nash equilibria, whereas strategy C is not a Nash equilibrium. By the results of \cite{nunezneves}, we know that for the value of $q$ used in (\ref{payofffullevo}) there is no interior equilibrium for the dynamics and no equilibrium on side $AC$. On the side $AC$ A is always fitter than C. On the side $AB$ we have an equilibrium, depicted as $P_{12}$ in Fig. \ref{fullevodet}, such that above $P_{12}$ on that side, B is fitter than A, but A is fitter than B below $P_{12}$. On the side $BC$ we have an equilibrium $P_{23}$ such that B is fitter than C above it and C is fitter than B below it.

We also show in Fig. \ref{fullevodet} the lines in which the deterministic fitnesses $F$, $G$ and $H$ defined in (\ref{detfitnesses}) are pairwise equal. These lines divide the $ABC$ triangle in regions where the fitness ranking is fixed. It can be seen that above the blue line we have $G>H>F$, between the blue and red lines we have $H>G>F$, between the red and green lines we have $H>F>G$ and, finally, below the green line we have $F>H>G$.

\begin{figure}
	\begin{center}
		\includegraphics[width= 0.9\textwidth]{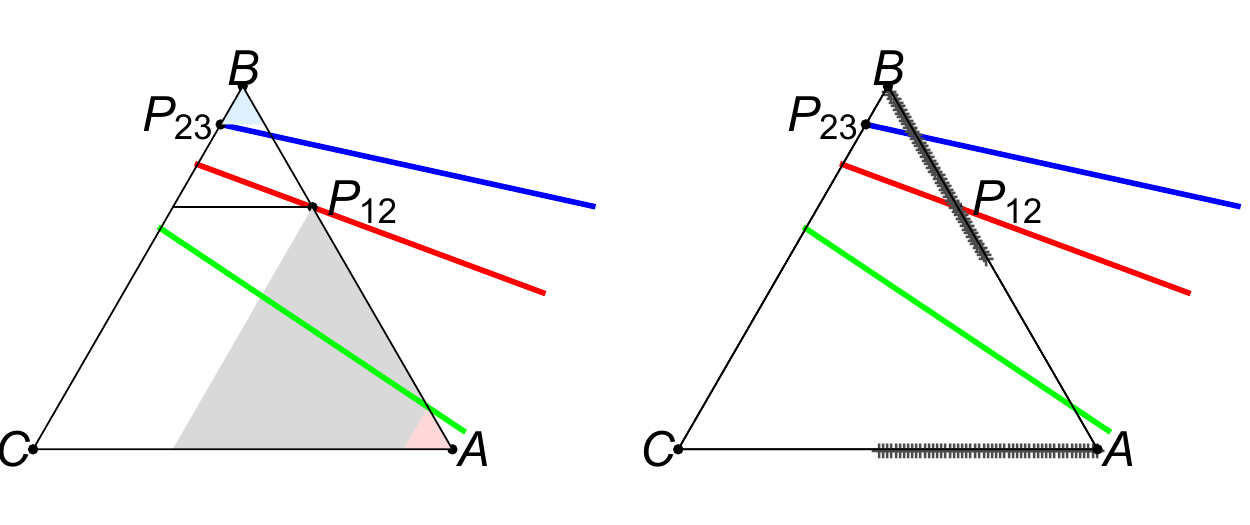}
		\caption{The red, green and blue lines and points $P_{12}$ and $P_{23}$ are the same as defined in the caption of Fig. \ref{fullevodet}. Left panel: The red region close to vertex A is the one in which we can use Theorem \ref{theoAstrictNash} to prove that $A_N(x,y)$ tends to 1 when $N \rightarrow \infty$. The blue region close to B is, by an analogue of the same theorem, the one in which we can prove that $B_N(x,y)$ tends to 1 when $N \rightarrow \infty$. Arguments in this section prove that besides the red region we can prove that $\lim_{N \rightarrow \infty} A_N(x,y)=1$ also in the gray region. Right panel: The crosses on sides $AB$ and $AC$ are the numerically determined locations, for each fixed value of $i$, of the minimum over $j$ of the $Z^+_{ij}$. The pay-off matrix is (\ref{payofffullevo}) and population size is $N=200$.
			\label{regions}}
	\end{center}
\end{figure}

Let $x^*_1$ denote the fraction of A individuals at the point in which the line $F=H$ intercepts the $AB$ side. The fitness ranking given above shows that if $x>x^*_1$, then $F>H>G$. We can then readily apply Theorem \ref{theoAstrictNash} to conclude that for initial conditions $(x,y)$ in the red region close to vertex A in the left panel of Fig. \ref{regions} we have $\lim_{N \rightarrow \infty} A_N(x,y)=1$. Similarly, if $y^*_1$ denotes the fraction of B individuals at the point $P_{23}$, then, for $y>y^*_1$ we have $G>H>F$. By an analogue of the same Theorem, $\lim_{N \rightarrow \infty} B_N(x,y)=1$ if $(x,y)$ is in the blue region in the left panel of Figure \ref{regions}.

Another of our results we can use is Theorem \ref{partialexttheo}. If we denote $y^*_2$ the fraction of type B individuals at point $P_{12}$, then the fitness ranking already exhibited shows that B is the least fit strategy for all $(x,y) \in \Lambda$ with $y<y^*_2$. By an analogue of Theorem \ref{partialexttheo}, $B_N(x,y) \stackrel{N \rightarrow \infty}{\rightarrow} 0$ for such points.

Although $B_N(x,y) \stackrel{N \rightarrow \infty}{\rightarrow} 0$ does not mean that the fixation probability for strategy A should be large, the above result along with the orbits in Fig. \ref{fullevodet} suggest that we might possibly find a region larger than the red region in the left panel of Fig. \ref{regions} in which $A_N(x,y) \stackrel{N \rightarrow \infty}{\rightarrow}1$.

As already commented at the beginning of Sect. \ref{secnash}, the results proved there, Theorem \ref{theoAstrictNash} included, did not rely on locating the maximum or minimum over $j$ of the $Z^+_{ij}$. We will exhibit an argument based on simple numerical calculations and also on Theorem \ref{theogeneralbounds} that strongly suggests that we can find a larger region in which $A_N(x,y) \stackrel{N \rightarrow \infty}{\rightarrow}1$. In order to use Theorem \ref{theogeneralbounds}, we will need to locate the cited minima.

A problem here is that $Z^+_{ij}$ is a complicated function depending on $i, j, w, N$ and the pay-off matrix elements. We were not able to rigorously locate for each $i$ the maximum over $j$ of the $Z^+_{ij}$. Even if we simplified the expression for $Z^+$ by taking a deterministic limit, as we have done in many places in this paper, we still could not show that the location of the minima of the expression was stable under small variations of the pay-off matrix values. Instead of presenting such long and not so conclusive calculation, we preferred to fix the pay-off matrix (\ref{payofffullevo}), take population size $N=200$ and numerically locate for each fixed value of $i$, the minimum over $j$ of the $Z^+_{ij}$. This is a very simple computational task and the results are shown at the right panel in Fig. \ref{regions}. We repeated the same task for larger values of $N$ and the results were not changed.

If we take for exact the result of these numerical calculations, we can now enlarge the region in which we have $A_N(x,y) \stackrel{N \rightarrow \infty}{\rightarrow}1$. Let $x^*_2$ be the fraction of A individuals at point $P_{12}$ and $x^*_3$ be the same for the point in the right panel of Fig. \ref{regions} in which the location of the minima of $Z^+_{ij}$ changes abruptly from the $AB$ side to the $AC$ side. 

Reminding the reader of Proposition \ref{propminmaxz} and using the notation of Theorem \ref{theogeneralbounds}, we see that for $[Nx^*_2]<i < [Nx^*_3]$ we have 
\[a^{low}_i = Z^+_{i,N-i}= \frac{i (N-i)f_{i,N-i}}{i f_{i,N-i}+(N-i) g_{i,N-i}}\]
and
\[ b^{low}_i = Z^-_{i,N-i}= \frac{i (N-i)g_{i,N-i}}{i f_{i,N-i}+(N-i) g_{i,N-i}}\;.\] 
Thus, for these values of $i$, $r^{low}_i= a^{low}_i/b^{low}_i= f_{i,N-i}/g_{i,N-i}$. For large enough $N$, $f_{i,N-i}/g_{i,N-i}$ is arbitrarily approximated by $F(\frac{i}{N}, 1-\frac{i}{N})/ G(\frac{i}{N}, 1-\frac{i}{N})$. As $F>G$ at all points on side $AB$ with $x>x^*_2$, then $r^{low}_i>1$ if $[Nx^*_2]<i < [Nx^*_3]$ and $N$ is large enough.

If, on the other hand, $i> [Nx^*_3]$, then 
\[a^{low}_i = Z^+_{i,0}= \frac{i (N-i)f_{i,0}}{i f_{i,0}+(N-i) h_{i,0}}\]
and
\[ b^{low}_i = Z^-_{i,0}= \frac{i (N-i)h_{i,0}}{i f_{i,0}+(N-i) h_{i,0}}\;.\] 
It follows that if $i > [Nx^*_3]$, then $r^{low}_i= f_{i,0}/h_{i,0}>1$, because on the $AC$ side $F>H$ everywhere.

Because $r^{low}_i$ is determined by relative fitnesses calculated on sides $AC$ and $AB$ and on these sides we are dealing with Moran processes with two strategies, it is easy to prove a bit more: if we accept that the numerically calculated location of the minima of $Z^+_{ij}$ for each $i$ is exact, then there exists $s>1$ such that for $N$ large enough and $i>N x^*_2$, $r^{low}_i> s$. Using Theorem \ref{partialfixationtheo}, we may conclude that for all $(x,y) \in \Lambda$ with $x>x^*_2$, $A_N(x,y) \stackrel{N \rightarrow \infty}{\rightarrow}1$. This region is the union of the gray and red regions depicted in the left panel of Fig. \ref{regions}.

\section{Conclusions} \label{secconc}
When we started working on the subject of Moran process with three or more strategies, we had as an optimistic goal to provide a complete classification of all possible behaviors, either as \cite{taylor} did for the Moran process with two strategies, or as \cite{Bomze83} did for the replicator dynamics with three strategies.
We see we are still very far from achieving this goal, but we believe that this paper may be a good starting point for further work. 

As a first important achievement, we introduced coupling of stochastic processes as a tool for obtaining results for the Moran process with three or more strategies. More specifically, we used coupling to obtain upper and lower bounds of general validity for the fixation probabilities. 

We do not claim that our general upper and lower bounds in Theorem \ref{theogeneralbounds} are optimal. In fact, although Theorems \ref{lowboundtheo} and \ref{upperboundtheo} allow more flexibility, the general recipe of Theorem \ref{theogeneralbounds} with its maxima and minima taken over all $j$ may produce too small lower bounds or too large upper bounds. Nonetheless, this general recipe has proved powerful enough for proving in some cases, see results in Sect. \ref{secnash}, that some strategies may fixate or be extinct with large probability if the population size $N$ is large.

In Sect. \ref{seccoop} we applied these results to a concrete problem. This application showed at the same time usefulness and weakness of the results in the preceding sections. Usefulness because we readily found some regions in which some strategies were either fixated with large probability, or extinct with large probability. Weakness because when we tried to find a larger region in which the G strategy had a large fixation probability, we were faced with the difficulty in determining, for fixed $i$,  the location of the points in which $Z^+_{ij}$ was minimized over $j$.

We hope that our results may prove useful in other applications, or else, that work in other applications may suggest some better bounds for fixation probabilities.

\appendix
\section{Some results on birth-death processes}  \label{appbdprocesses}
This appendix collects some results which we did not want to insert in the main text of the paper, because they have to do only with birth-death processes. Despite that, these results were used in the proof of the theorems in Section \ref{secnash}, all of them referring to Moran processes with three strategies. The more important results here are Theorems \ref{partialfixationtheo} and \ref{partialfixationtheoanalogo}, which are cited in the proofs in the main text. The propositions which precede them are necessary for their proofs.

The first result here deals with comparing fixation probabilities for two birth-death processes in which the birth to death ratio is larger in one process than in the other. We observe that this result might be proved by a coupling argument similar to the one shown in Theorem \ref{lowboundtheo}. We opt here for a direct proof using the exact expressions (\ref{piiformula}) for the fixation probabilities.
\begin{proposition}  \label{BDcomparisonprop}
	Consider two birth-death processes with the same set of states $S=\{0,1,2, \dots, N\}$. Let $r_i\equiv a_i/b_i$ be the birth to death ratio in the first process and $s_i\equiv a'_i/b'_i$ be the ratio in the second process. Let also $\pi_i$ and $\pi'_i$ denote the respective fixation probabilities in state $N$. If $r_i >s_i$ for $i=1,2, \dots, N-1$, then $\pi_i>\pi'_i$ for all $i \in S \setminus\{0,N\}$.
\end{proposition}

\textbf{Proof}
	We start by rewriting expression (\ref{piiformula}) for $\pi_i$:
	\begin{align*}
		\pi_i &=\, \frac{1+ \sum_{j=1}^{i-1} \prod_{k=1}^j r_k^{-1}}{1+ \sum_{j=1}^{N-1} \prod_{k=1}^j r_k^{-1}}\,=\, \frac{1+ \sum_{j=1}^{i-1} \prod_{k=1}^j r_k^{-1}}{1+ \sum_{j=1}^{i-1} \prod_{k=1}^j r_k^{-1}+ \sum_{j=i}^{N-1} \prod_{k=1}^j r_k^{-1}}\\
		&= \; \frac{1}{1+ \frac{\sum_{j=i}^{N-1} \prod_{k=1}^j r_k^{-1}}{1+ \sum_{j=1}^{i-1} \prod_{k=1}^j r_k^{-1}}}\;.
	\end{align*}
	A similar expression may be written for $\pi'_i$ just by writing $s_k$ in place of $r_k$.
	Let
	\[d_i= \frac{\sum_{j=i}^{N-1} \prod_{k=1}^j r_k^{-1}}{1+ \sum_{j=1}^{i-1} \prod_{k=1}^j r_k^{-1}}\]
	be the denominator minus 1 in the last expression and $d'_i$ be the same expression with $r_k$ exchanged by $s_k$. We will prove that $d'_i-d_i>0$, which of course implies $\pi_i>\pi'_i$.
	
	\begin{align*}
		&d'_i-d_i=\\ 
		&\frac{\left(\sum_{j=i}^{N-1} \prod_{k=1}^j s_k^{-1}\right)\left(1+ \sum_{j=1}^{i-1} \prod_{k=1}^j r_k^{-1}\right)-\left(\sum_{j=i}^{N-1}\prod_{k=1}^j r_k^{-1}\right)\left(1+ \sum_{j=1}^{i-1} \prod_{k=1}^j s_k^{-1}\right)}{\left(1+ \sum_{j=1}^{i-1} \prod_{k=1}^j r_k^{-1}\right)\left(1+ \sum_{j=1}^{i-1} \prod_{k=1}^j s_k^{-1}\right)}\\
		&= \frac{\sum_{j=i}^{N-1}\left(\prod_{k=1}^j s_k^{-1}- \prod_{k=1}^j r_k^{-1}\right)}{\left(1+ \sum_{j=1}^{i-1} \prod_{k=1}^j r_k^{-1}\right)\left(1+ \sum_{j=1}^{i-1} \prod_{k=1}^j s_k^{-1}\right)}\\&+\frac{\left(\sum_{j=i}^{N-1} \prod_{k=1}^j s_k^{-1}\right)\left(\sum_{j=1}^{i-1} \prod_{k=1}^j r_k^{-1}\right)-
			\left(\sum_{j=i}^{N-1} \prod_{k=1}^j r_k^{-1}\right)\left(\sum_{j=1}^{i-1} \prod_{k=1}^j s_k^{-1}\right) }{\left(1+ \sum_{j=1}^{i-1} \prod_{k=1}^j r_k^{-1}\right)\left(1+ \sum_{j=1}^{i-1} \prod_{k=1}^j s_k^{-1}\right)}\;.
	\end{align*}
	Using the fact that $s_k^{-1}>r_k^{-1}>0$ for all $k$, both numerator and denominator in the first term in the last expression are clearly positive. To see that the second term is positive, too, notice that its denominator is the same of the first term, and its numerator, with some patience, may be rewritten as
	\[\sum_{j=1}^{i-1} \left(\prod_{k=1}^j r_k^{-1} s_k^{-1}\right)\, \sum_{\ell=1}^{N-i} \left(\prod_{m=j+1}^{N-\ell} s_m^{-1}- \prod_{m=j+1}^{N-\ell} r_m^{-1}\right) \;,\]
	now manifestly positive.
$\blacksquare$

\begin{proposition}\label{partialfixationprop}
	Consider a birth-death process with state space $S=\{0,1,2, \dots, N\}$. If $r_i$ is the ratio of birth to death probabilities and $i^{\ast} \in \{1, 2, \dots, N-1\}$ is some fixed state, then the probability $\pi_{i \backslash i^{\ast}}$ that the process starts at $i>i^{\ast}$ and fixates at state $N$ without ever passing by state $i^{\ast}$ is 
	\begin{equation} \label{piii*}
		\pi_{i \backslash i^{*}} = \frac{1 + \sum_{\ell=1}^{i-i^{*}-1}\prod_{k=1}^{\ell}r_{i^{*}+k}^{-1}}{1 + \sum_{\ell=1}^{N-i^{*}-1}\prod_{k=1}^{\ell}r_{i^{*}+k}^{-1}}\;. 
	\end{equation}
\end{proposition}

\textbf{Proof}
	Our result (\ref{piii*}) may be obtained from (\ref{piiformula}), noticing that the boundary condition $\pi_0=0$ is replaced by $\pi_{i^{\ast} \backslash i^{\ast}}=0$ and the set of states $\{0,1,2, \dots, N\}$ is replaced by $\{i^{\ast}, i^{\ast}+1, \dots, N\}$.
	$\blacksquare$

The following result, Proposition \ref{BDpartialcomparisonprop}, is just a straightforward adaptation of the result in Proposition \ref{BDcomparisonprop} to the fixation probability $\pi_{i^{\ast} \backslash i^{\ast}}$ defined in Proposition \ref{partialfixationprop}. As the proof is a mere repetition, we do not write it here.

\begin{proposition}  \label{BDpartialcomparisonprop}
	Consider two birth-death processes with the same set of states $S=\{0,1,2, \dots, N\}$. Let $r_i\equiv a_i/b_i$ and $s_i\equiv a'_i/b'_i$ be the respective birth to death ratios. Suppose that there exists $i^*$ such that $r_i >s_i$ for $i=i^*+1,i^*+2, \dots, N-1$. If $i>i^*$ and $\pi_{i \backslash i^{*}}$ and $\pi'_{i \backslash i^{*}}$ denote the fixation probabilities in state $N$ with the additional condition that the process never passes by state $i^*$, then $\pi_{i \backslash i^{*}}>\pi'_{i \backslash i^{*}}$ for all $i \in \{i^*+1,i^*+2, \dots, N-1\}$.
\end{proposition}

The next result is the key ingredient in the proof of Theorem \ref{theobeforeAstrictNash} in the main text.
\begin{theorem}\label{partialfixationtheo}
	Suppose that for large enough values of $N$ we have a family of birth-death processes with birth to death ratios $r_i^{(N)}$ and fixation probabilities $\pi_i^{(N)}$, $i=1, 2, \dots, N-1$. Let $x\in (0,1)$ and $\Pi_N(x) \equiv \pi_{[Nx]}^{(N)}$. If there exist $s>1$ and $x^* \in (0,1)$ such that
	for $N$ large enough and $i>N x^*$ we have $r_i^{(N)}>s$, then 
	\[\lim_{N \rightarrow \infty}\Pi_N(x)=1 \] 
	for $x>x^*$. 
\end{theorem}

\textbf{Proof}
	Let $\pi'_i$ be the fixation probability of a birth-death process with frequency independent fitness $s_i=s$ and let $\pi'_{i\setminus i^*}$ be as in Proposition \ref{BDpartialcomparisonprop}. Summing the geometric progressions appearing in (\ref{piii*}) when $r_i$ is replaced by $s$, we get
	\[ \pi'_{i\setminus i^*} \,=\,  \frac{1-s^{-(i-i^{*})}}{1-s^{-(N-i^{*})}}\;.\]
	Suppose $x>x^*$ and $N$ large enough so that $[Nx]>[Nx^*]$. Of course, $\pi_{[Nx]}^{(N)} \geq \pi_{[Nx]\setminus[Nx^*]}^{(N)}$. As, by Proposition \ref{BDpartialcomparisonprop}, we have $\pi_{[Nx]\setminus[Nx^*]}^{(N)} > \pi'_{[Nx]\setminus [Nx^*]}$, then
	\[\pi_{[Nx]}^{(N)}  >\frac{1-s^{-([Nx]-[Nx^{*}])}}{1-s^{-(N-[Nx^{*}])}}\;. \]
	Our conclusion follows because, if $s>1$, the last expression tends to 1 when $N \rightarrow \infty$. $\blacksquare$

The next result here is quite analogous to Theorem \ref{partialfixationtheo} in its proof, but it comes with a surprise: although we will be able to prove that, under the stated hypotheses, $\pi_{[Nx]\setminus[Nx^*]}^{(N)}$ tends to 0 as $N \rightarrow \infty$, we cannot conclude that $\pi_{[Nx]}^{(N)}$ tends to 0, too.
\begin{theorem}\label{partialfixationtheoanalogo}
	Suppose that for large enough values of $N$ we have a family of birth-death processes with birth to death ratios $r_i^{(N)}$ and fixation probabilities $\pi_i^{(N)}$, $i=1, 2, \dots, N-1$. Suppose also that there exist $0<s<1$ and $x^* \in (0,1)$ such that
	for all $N$ and $i>N x^*$ we have $r_i^{(N)}<s$. If $x>x^*$ and $\Pi_{N\backslash x^{*}}(x) \equiv \pi_{[Nx]\backslash [Nx^{*}]}^{(N)}$, then 
	\[\lim_{N \rightarrow \infty}\Pi_{N\backslash x^{*}}(x)=0\;. \] 
\end{theorem}

The proof of the above result is analogous to the proof of Theorem \ref{partialfixationtheo} and is left to the interested reader. We comment instead on why we cannot arrive at a result completely analogous to Theorem \ref{partialfixationtheo}.

The first reason is that inequality $\pi_{[Nx]}^{(N)} \geq \pi_{[Nx]\setminus[Nx^*]}^{(N)}$ used in proving Theorem \ref{partialfixationtheo} is still valid and we cannot in general conclude that a quantity larger than or equal to something tending to 0 tends to 0, too. 

More than that, we know that in a birth-death process for two strategies we can fulfill the hypotheses of Theorem \ref{partialfixationtheoanalogo} and still have $\lim_{N \rightarrow \infty} \pi_{[Nx]}^{(N)} =1$. This is proved for Moran processes with two strategies in \cite{graphshapes}, Theorem 5, if certain conditions on the pay-off matrix are valid. The conditions are $m_{11}<m_{21}$, $m_{12}>m_{22}$, i.e. neither of the two strategies is a Nash equilibrium, and
\begin{equation}
	\label{L(1)<0}
	L(1) \,\equiv  \, - \, \int_{0}^{1} \log \frac{1-w+w[m_{11}t+m_{12}(1-t)]}{1-w+w[m_{21}t+m_{22}(1-t)]} \,dt <0 \;.
\end{equation}

In the interesting situation in which the hypotheses of Theorem \ref{partialfixationtheoanalogo} are fulfilled and we also have $\lim_{N \rightarrow \infty} \pi_{[Nx]}^{(N)} =1$, we have for large $N$ and $x>x^*$ both $\Pi_N(x)$ close to 1 and $\Pi_{N\backslash x^{*}}(x)$ close to 0. This means that although fixation at state $N$ is very probable, most probably the chain will pass at least once (thus, it will probably pass many times) by $x^*$ before fixation occurs. 
An application of the above phenomenon, in which a repeller strategy in the replicator dynamics fixates with high probability in the Moran process, is given by \cite{McLooneetal}.

\section*{Acknowledgements}
This study was financed in part by the Coordena\c{c}\~ao de
Aperfei\c{c}oamento de Pessoal de N\'ivel Superior - Brasil (CAPES) -
Finance Code 001.


\begin{thebibliography}{10}
	
	\bibitem{allen}
	Linda J.~S. Allen.
	\newblock {\em An introduction to stochastic processes with applications to
		biology}.
	\newblock Chapman \& Hall/CRC, Boca Raton, FL, 2011.
	
	\bibitem{AntalScheuring}
	Tibor Antal and Istv{\'a}n Scheuring.
	\newblock Fixation of strategies for an evolutionary game in finite
	populations.
	\newblock {\em B. Math. Biol.}, 68(8):1923--1944, 2006.
	
	\bibitem{Bomze83}
	I.~M. Bomze.
	\newblock Lotka-{Volterra} and replicator dynamics: A two dimensional
	classification.
	\newblock {\em Biol. Cybernetics}, 48:201--211, 1983.
	
	\bibitem{ChalubSouza2016}
	Fabio A. C.~C. Chalub and Max~O. Souza.
	\newblock Fixation in large populations: a continuous view of a discrete
	problem.
	\newblock {\em J. Math. Biol.}, 72(1):283--330, 2016.
	
	\bibitem{graphshapes}
	Evandro~P. de~Souza, Eliza~M. Ferreira, and Armando G.~M. Neves.
	\newblock Fixation probabilities for the {Moran} process in evolutionary games
	with two strategies: graph shapes and large population asymptotics.
	\newblock {\em J. Math. Biol.}, 2018.
	
	\bibitem{hollander}
	Frank den Hollander.
	\newblock {\em Probability theory: the coupling method}.
	\newblock available at
	http://websites.math.leidenuniv.nl/probability/lecturenotes/CouplingLectures.pdf,
	accessed in June 8, 2018., 2012.
	
	\bibitem{DurandLessard}
	Guillermo Durand and Sabin Lessard.
	\newblock Fixation probability in a two-locus intersexual selection model.
	\newblock {\em Theor. Popul. Biol.}, 109:75 -- 87, 2016.
	
	\bibitem{ewens}
	Warren~J. Ewens.
	\newblock {\em Mathematical population genetics. {I}. , Theoretical
		introduction}.
	\newblock Interdisciplinary applied mathematics. Springer, New York, 2004.
	
	\bibitem{hofbauersigmund}
	J.~Hofbauer and K.~Sigmund.
	\newblock {\em Evolutionary Games and Population Dynamics}.
	\newblock Cambridge University Press, Cambridge, 1998.
	
	\bibitem{MaynardSmithPrice}
	J.~{Maynard Smith} and G.~Price.
	\newblock The logic of animal conflicts.
	\newblock {\em Nature}, 246:15 -- 18, 1973.
	
	\bibitem{McLooneetal}
	Brian McLoone, Wai-Tong~Louis Fan, Adam Pham, Rory Smead, and Laurence Loewe.
	\newblock Stochasticity, selection, and the evolution of cooperation in a
	two-level {Moran} model of the snowdrift game.
	\newblock {\em Complexity}, 2018.
	
	\bibitem{moran}
	P.~A.~P. {Moran}.
	\newblock {Random processes in genetics}.
	\newblock {\em P. Camb. Philos. Soc.}, 54(1):60, 1958.
	
	\bibitem{nowakbook}
	M.~Nowak.
	\newblock {\em Evolutionary Dynamics}.
	\newblock The Belknap Press of Harvard University Press, 1 edition, 2006.
	
	\bibitem{nowaksigmundnature}
	M.~A. Nowak and K.~Sigmund.
	\newblock Tit for tat in heterogeneus populations.
	\newblock {\em Nature}, 355:255--253, 1992.
	
	\bibitem{nowaknature}
	Martin~A. Nowak, Akira Sasaki, Christine Taylor, and Drew Fudenberg.
	\newblock Emergence of cooperation and evolutionary stability in finite
	populations.
	\newblock {\em Nature}, 428(6983):646--650, 2004.
	
	\bibitem{nunezneves}
	Irene N{\'u}{\~{n}}ez~Rodr{\'i}guez and Armando G.~M. Neves.
	\newblock Evolution of cooperation in a particular case of the infinitely
	repeated prisoner's dilemma with three strategies.
	\newblock {\em J. Math. Biol.}, 73(6):1665--1690, 2016.
	
	\bibitem{petrovsky}
	I.~G. Petrovsky.
	\newblock {\em Lectures on partial differential equations}.
	\newblock Dover Publications, 1992.
	
	\bibitem{taylor}
	Christine Taylor, Drew Fudenberg, Akira Sasaki, and Martin~A. Nowak.
	\newblock Evolutionary game dynamics in finite populations.
	\newblock {\em B. Math. Biol.}, 66(6):1621--1644, 2004.
	
	\bibitem{taylorjonker}
	P.~D. Taylor and L.~B. Jonker.
	\newblock Evolutionary stable strategies and game dynamics.
	\newblock {\em Math. Biosci.}, 40:145--156, 1978.
	
	\bibitem{wang2007evolutionary}
	Jing Wang, Feng Fu, Long Wang, and Guangming Xie.
	\newblock Evolutionary game dynamics with three strategies in finite
	populations.
	\newblock {\em arXiv preprint physics/0701315}, 2007.
	
\end{thebibliography}

\end{document}